\newif\ifShowKeys
\ShowKeystrue
\documentclass[11pt,a4paper]{article} 
\pdfoutput=1
\usepackage[no-natbib-sort]{my-jheppub}
\usepackage{amsmath, amssymb}

\usepackage{mathpazo}
\usepackage{bm}
\usepackage{environ}
\usepackage{mathrsfs}
\usepackage{array,arydshln}
\usepackage{booktabs}

\usepackage{graphicx,epsfig}
\usepackage{epic}

\usepackage{tensor} 							% Ratcliffe package to write tensors

\usepackage{float}
\usepackage[dvipsnames]{xcolor}
\definecolor{maroon}{rgb}{0.8,0.3,0.}

\usepackage{slashed}
\usepackage[nodayofweek]{date time}

\usepackage{hyperref}

\usepackage{aurical}
\usepackage[T1]{fontenc}

\usepackage{framed}
\definecolor{shadecolor}{RGB}{255, 230, 204}

\allowdisplaybreaks

\newmuskip\pFqmuskip

\newcommand*\pFq[6][8]{%
  \begingroup % only local assignments
  \pFqmuskip=#1mu\relax
  \mathcode`\,=\string"8000
  \begingroup\lccode`\~=`\,
  \lowercase{\endgroup\let~}\pFqcomma
  {}_{#2}F_{#3}{\left[\genfrac..{0pt}{}{#4}{#5};#6\right]}%
  \endgroup
}

\newcommand*\pFtildeq[6][8]{%
  \begingroup % only local assignments
  \pFqmuskip=#1mu\relax
  \mathcode`\,=\string"8000
  \begingroup\lccode`\~=`\,
  \lowercase{\endgroup\let~}\pFqcomma
  {}_{#2}\widetilde{F}_{#3}{\left[\genfrac..{0pt}{}{#4}{#5};#6\right]}%
  \endgroup
}

\newcommand{\pFqcomma}{\mskip\pFqmuskip}

\newcommand{\be}{\begin{equation}}
\newcommand{\ee}{\end{equation}}

\newcommand{\mc}{\mathcal }

\newcommand{\bv}{\begin{verbatim}}
\newcommand{\ev}{\end{verbatim}}

\newcommand{\la}{\label}

\newcommand{\eps}{\varepsilon}

\newcommand{\vp}{\varphi}

\def \ov {\over}
\def \ci {\cite}
\def \foot {\footnote}

\def \b{\beta}
\def \m {\mu}
\def \n {\nu}
\def \del{\partial}

\def \ep{\epsilon}
\newcommand{\rf}[1]{(\ref{#1})}
\def \r {\rho}
\def \k {\kappa}
\def \l {\lambda}
\def \iffa {\iffalse}
\def \d {\partial} 
\def \a  {\alpha}

\def \cc {{\rm c}}
\def \aa {{\rm a} }
\def  \ba { \begin{align} }
\def  \ea { \end{align} }
\def \rp {{\rm p}}
  \def \tr  {{\rm tr }}   
\def \na {\nabla}
 
\def \ha {\tfrac{1}{2}}   \def \H  {{\mathbb H}}

\newcommand{\renyi}{{R\'enyi} }

\title{$C_T$   for  higher derivative conformal fields
and  anomalies of $(1,0)$   superconformal  6d theories}

\author[a]{Matteo Beccaria} 
\author[,b,c]{ and \ \ Arkady A. Tseytlin\footnote{Also at Lebedev Institute, Moscow.}}

\abstract{
\ 
In  arXiv:1510.02685  we proposed the  linear relations  between the   Weyl anomaly 
$\cc_1,\cc_2,\cc_3$ coefficients   and the 4 coefficients in the chiral anomaly polynomial  for (1,0)   superconformal  6d  theories.  These   relations   were  determined  up to  one free  parameter $\xi$   and  its   value was then   conjectured   using    some additional assumptions. 
A  different  value for $\xi$ was  recently    suggested  in   arXiv:1702.03518 using an  alternative  method. 
 Here we confirm  that this   latter     value  is  indeed the correct  one 
   by providing   an additional  data point:  the 
Weyl anomaly coefficient $\cc_3$  for  the   higher derivative   (1,0)
superconformal  6d  vector multiplet. This multiplet   contains  the 
4-derivative  conformal gauge vector, 3-derivative   fermion and  2-derivative  scalar. 
We find the  corresponding  value of  $\cc_3$  which is proportional to the  coefficient $C_T$ 
in the 2-point function of stress tensor 
using   its   relation to  the first derivative of the Renyi entropy   or 
the second derivative of 
the free  energy on  the product of thermal circle and 5d hyperbolic space. % $S^1_q \times \H^5$.  
We   present some general results   of the computation of   the \renyi entropy     and    $C_T$  
from the partition function on $S^1 \times \H^{d-1}$
 for  higher derivative conformal  scalars, spinors  and vectors   in  even dimensions.
 We also give an independent   derivation  of the conformal anomaly  coefficients of the 6d  higher derivative vector multiplet 
 from the Seeley-DeWitt coefficients  on an Einstein  background. 
}

\affiliation[a]{Dipartimento di Matematica e Fisica Ennio De Giorgi,\\
Universit\`a del Salento \& INFN, Via Arnesano, 73100 Lecce, 
Italy}

\affiliation[b]{Kavli Institute for Theoretical Physics,  University of California, 
Santa Barbara, CA 93106, USA}
    
    \affiliation[c]{The Blackett Laboratory, Imperial College, London SW7 2AZ, U.K.}
                   
\emailAdd{matteo.beccaria@le.infn.it} \emailAdd{tseytlin@imperial.ac.uk}

\begin{document}

\def \g  {\gamma}\def \te {\textstyle} 
\def \zp {z_\Phi}

%\date{\currenttime}

%\begin{flushleft}\boxed{\small{\tt \today \ \ - \ \  \currenttime }}\end{flushleft}

\begin{flushright}\small{Imperial-TP-AT-2017-05}, \ \ \ \  \small{NSF-KITP-17-061}\ \end{flushright}	
%\vskip -0.2pt			% report number
%\begin{flushright}\small{NSF-KITP-17-061}\end{flushright}	

\def \SS {{\mathbb S}}
\def \Ss {{S}}  \def \V  {{\rm V}}

\maketitle
\flushbottom

  \def\no{\nonumber}
\def \const {{\rm const}}  \def \r {\rho}\def \s {\sigma}

\def \LI  {{\Lambda_{\rm IR}}}
\def \LU  {{\Lambda_{\rm UV}}}
\def \dd {{d \ov 2}}
\def \VV  {\overline{\V}}

\section{Introduction}

\def \xibt {\xi_{_{{\rm BT}}}}
\def \xiyz {\xi_{_{{\rm YZ}}}}

The conformal anomaly of a classically  Weyl invariant theory in 6d   depends 
on 4 independent coefficients  $\aa,\cc_1,\cc_{2},\cc_3$  
\cite{Bonora:1985cq,Deser:1993yx,Bastianelli:2000hi,Boulanger:2007ab}
\be
\label{1.1}
  (4\pi)^{3} \langle T^\m_\m\rangle = 
  -\aa\,E_{6}+ \cc_{1}\,I_{1}+\cc_{2}\,I_{2}+\cc_{3}\,I_{3}\   ,  
\ee
where  %$\mc A_6$  stands for the  integrand  of the Seeley coefficient $b_6$, 
% MB-1 : added sketchy definition of E_{6} 
$E_{6} $ is the  6d Euler density 
   and  the three   Weyl invariants  are  
$I_1  = C_{\a\m\n\b}C^{\m\r\s\n}C_{\r\a\b\s}$, \ \  $I_2 = C_{\a\b\m\n}C^{\m\n\r\s}C_{\r\s\a\b},$\ \ $ I_3 = C_{\m\n\a\b} \na^2 C^{\m\n\a\b} + ... $.
As  \rf{1.1}   appears in  the  log UV divergent part of  the effective action, 
  $\cc_3$    can be  determined  from  the   2-point function of stress tensor $ \langle T T\rangle $, 
  $\cc_2$ and $\cc_3$ --  from the 3-point function\footnote{ $\langle TTT\rangle$   in 6d depends on  three  parameters, but one of them is related to the 2-point function or $\cc_3$ 
   by a conformal 
Ward identity \cite{Osborn:1993cr,Erdmenger:1996yc,Bastianelli:1999ab}.}   and the $\aa$-coefficient --   
from the  4-point function. 

%see  for instance \cite{Bastianelli:2000hi}.
%Thus, in 6d there are 4 independent conformal anomaly coefficients.
% \footnote{
%Instead, on a Ricci flat background  one has the identities $E_{6} = 32\,(2\,I_{1}+I_{2})$ and 
%$I_{3} = 4\,I_{1}-I_{2}$, 
%so that \rf{1.1} reduces to  $
% A_{6}\big|_{R_{mn}=0}  =\te -\big[\aa - \frac{1}{192} (\cc_1 + 4 \cc_{2})\big]\, E_{6}
% +(\cc_{1}-2\cc_{2}+6\cc_{3})\,I_{1}$.
% }
In the presence of   $(1,0)$  supersymmetry  one expects that the   Weyl invariants $I_i$   are 
bosonic parts of only two possible 6d superinvariants, {\em i.e.}  the coefficients  $\cc_i$   should  satisfy  one linear relation. 
As  discussed in \cite{Beccaria:2015ypa}, free-theory  calculations \cite{Bastianelli:2000hi},  
strong-coupling  (holography) arguments   \ci{Kulaxizi:2009pz}, and studies in other contexts  
\cite{Hofman:2008ar,Safdi:2012sn,Bueno:2015lza}  indicate   that 
this  relation  is\footnote{In the case of  $(2,0)$    supersymmetry, 
the three invariants  $I_i$  are parts of a single  superinvariant  (6d conformal supergravity action \ci{Beccaria:2015uta,Butter:2016qkx,Butter:2017jqu})
  and thus  % conformal anomaly coefficients 
$\cc_i$ obey the  additional   constraint $\cc_1 - 4 \cc_2=0$. Then  
 there is only one independent $\cc$-coefficient:  
$\cc_1 =4 \cc_2 = -12 \cc_3 $.
This  relation holds  for the  free $(2,0)$   
tensor multiplet   \cite{Bastianelli:2000hi} as well as for 
the large $N$ strong coupling limit of the  interacting  $(2,0)$  theory  described by  
supergravity in AdS$_7$  \ci{Henningson:1998gx}. 
}
\be \la{1.2}
%(1,0): \qquad \qquad 
\cc_{3} =\te - {1\ov 6} ( \cc_1 - 2 \cc_2) \ . 
%\cc_1 -  2 \cc_2 + 6 \cc_3 =0.
\ee
  The  6d  chiral ($SU(2)$ R-symmetry and gravitational) 
  anomalies are encoded in the 8-form polynomial
  parametrized   by  4  numerical coefficients  $(\a,\b, \g, \delta)$.\foot{Explicitly, 
  % has the following general structure \ci{Frampton:1983ah,AlvarezGaume:1983ig,Zumino:1983rz,Faddeev:1985iz} 
 %  \ba 
 %  \la{1.3}
% &
$   \mc I_{8} =\te  \frac{1}{4!}\big(\alpha\,c_{2}^{2}+\beta\,c_{2}\,\rp_{1}+
\gamma\,\rp_{1}^{2} +\delta\,\rp_{2} \big) \ ,\ \ \ \ 
    c_2 = \tr\,  F^2 \ , \ \ \    \rp_1 = -\tfrac{1}{2}\,\tr\,  R^2, \ \ \  \te \rp_2 = -\frac{1}{4}\,\tr\,   R^4 +\frac{1}{8}\,(\tr\,R^{2})^{2}\ . $ %\la{1.4}
%\end{align}
}
The   6d  chiral and Weyl anomalies belong to  a supersymmetry 
multiplet \ci{Howe:1983fr,Manvelyan:2000ef,Manvelyan:2003gc} and 
as in the 4d  case % \cite{%Duff:1993wm,
\ci{Buchbinder:1986im,Anselmi:1997ys,Aharony:2007dj,Kuzenko:2013gva}
  one expects  to find  linear relations between their coefficients, 
  i.e.   between $(\aa,\cc_1,\cc_2,\cc_3)$ and  $(\a,\b, \g, \delta)$.
%where  $\vec \a = (\a,\b, \g, \delta)$  are numerical coefficients.
%\foot{The gauge bundle form
%$c_2$  should  not be confused with the conformal  anomaly coefficient $\cc_2$.}    
%v2
Ref. \cite{Cordova:2015fha} derived  such  relation for the  $\aa$-coefficient
using   supersymmetry  and the results  from the   background   supergravity couplings
%Guided by several explicit examples,  
%ref.  \cite{Cordova:2015fha} proposed  such  a  linear   relation for the  $\aa$-coefficient
\foot{In our normalization the a-anomaly of (2,0) tensor multiplet is 
 $\aa(T^{(2,0)})=- {7\ov 1152}$.}
 %valid in   $(1,0)$ superconformal 
%theories. It was found by determining the 4  parameters  $\vec k$  in the expected linear 
%relation $\aa= \vec k \cdot \vec \a$  
%from  several  explicit  examples and reads
\be
\label{1.5}
\aa = \te - {1 \ov 72} \big( \alpha-\beta+\gamma 
+\frac{3}{8}\,\delta\big)\ . \ee
The 4 coefficients  in \rf{1.5}   could  have been  be fixed also from the  
   $\aa$-anomalies  for the 4 multiplets:  free tensor, free hyper, the  (2,0)  multiplet 
at large $N$  {\it and}  the higher derivative vector multiplet  (the  $\aa$-anomaly of which was found    in 
\cite{Beccaria:2015uta,Beccaria:2015ypa} after    \cite{Cordova:2015fha}  already appeared).

%It is natural to expect a similar relation for the three $\cc_{i}$ anomaly coefficients.
 Assuming  that similar linear relations  exist  also for $\cc_1,\cc_2,\cc_3$, 
   in \cite{Beccaria:2015ypa}  we  attempted to fix their form   using the 
 available data about  $\cc$-anomalies  of  particular   (1,0)  superconformal  theories. 
 % and tensor $(2,0)$  6d supermultiplets (in (2,0)   case an extra input is that  linear relations should respect $\cc_1=4 \cc_2$). 
 The 
 linear relations for $\cc_1$  and $\cc_2$ in terms of     $(\a,\b, \g, \delta)$ 
   contain,  in general,  8   coefficients 
 ($\cc_3$ is    given by  \rf{1.2}). 
 We first  used  the values of anomaly coefficients  for   free   scalar $S^{(1,0)} $   and tensor 
 $T^{(1,0)} $  multiplets  
%These fields may be combined into $(1,0)$  scalar and tensor   and $(2,0)$  tensor   %supermultiplets
 %(here we indicate   chirality of the fields): 
 \be 
S^{(1,0)} = 4\varphi+2\psi^{-}, \ \ \qquad   T^{(1,0)} = \varphi+2\psi^{-}+T^{-}\ ,\ \ \qquad 
T^{(2,0)} = S^{(1,0)}+T^{(1,0)} %= 5\varphi+4\psi^{-}+T^{-} 
\  \la{1.6}\ee
 built out of  the standard 
 2-derivative real  scalar $\vp$, Majorana-Weyl (MW)  spinor $\psi$  and (anti) selfdual 
 rank 2 tensor  $T$  with  known   Weyl anomalies  \ci{Bastianelli:2000hi}.
%\end{align}
This  gave  4  coefficients out of 8.  One more coefficient   was 
 fixed   by considering 
  the 4-derivative vector  multiplet $V^{(1,0)}$  (see  \rf{1.10} below) 
  on a Ricci-flat background  when its Weyl anomalies  can be readily  computed. 
Two  more coefficients  were   found  from the known anomalies  of 
 interacting (2,0)  $A_N$  theory (see \ci{Beem:2014kka,Cordova:2015vwa} and refs. there).  
 As a result, we were able to  find the form  of 
 the relations for $\cc_{i}$  in terms of  $(\a,\b, \g, \delta)$ up to one undetermined    parameter  $\xi$, i.e.
   %. The result is 
\begin{align}
\la{1.7}
\cc_{1} &= -\tfrac{4}{3}\,\alpha+(\tfrac{146}{63}+\tfrac{6}{7}\,\xi)\,\beta+(
-\tfrac{80}{63}+\tfrac{4}{7}\,\xi)\,\gamma+\xi\,\delta\ , \notag \\
\cc_{2} &= -\tfrac{1}{3}\,\alpha+(\tfrac{11}{126}-\tfrac{3}{14}\,\xi)\,\beta
+(-\tfrac{22}{63}-\tfrac{1}{7}\,\xi)\,\gamma+(-\tfrac{1}{2}-\tfrac{1}{4}\,\xi)
\,\delta\ ,
% \\
%\cc_{3} &=\te - {1\ov 6} ( \cc_1 - 2 \cc_2) \ . \no
%+ \tfrac{1}{9}\,\alpha+(-\tfrac{5}{14}-\tfrac{3}{14}\,\xi)\,\beta+(\tfrac{2}{21}
%-\tfrac{1}{7}\,\xi)\,\gamma+(-\tfrac{1}{6}-\tfrac{1}{4}\,\xi)\,\delta. \notag
\end{align}
with $\cc_3$ given by \rf{1.2}.
We  then   conjectured that  the  value of    $\xi$   should  be 
%In \cite{Beccaria:2015ypa}, we proposed the special value 
\be
\la{1.8}
\xibt  = -\tfrac{31}{27} \ . 
\ee
% ending up with   \be \label{1.9}
%\begin{split}
%&\te  \cc_{1} = -\frac{4}{3}\,(\alpha-\beta)-\frac{52}{27}\,\gamma-\frac{31}{27}\,\delta\ , \qquad \qquad 
%\cc_{2} = -\frac{1}{3}\,(\alpha-\beta)-\frac{5}{27}\,\gamma-\frac{23}{108}\,\delta \ . 
%\end{split}
%\ee
This  particular  choice was 
%AT28
motivated  \cite{Beccaria:2015ypa}
by  certain special features in the rank dependence
of the c-anomaly of  particular  interacting $(1,0)$ superconformal theories   and also 
by  potential 
 relation  between 6d and  4d  anomalies  for  certain  (1, 0) theories compactified 
 on   2-torus  \cite{Ohmori:2015pua}. 

%leading to the specialized form of (\ref{1.7})
Recently,  the same  expressions for $\cc_i$  \rf{1.7},\rf{1.2}    but with  a different  value  of $\xi$
\be
\la{188}
\xiyz  = -\tfrac{8}{9} \ , 
\ee
 were found  in %using  different approach 
   \ci{Yankielowicz:2017xkf}   in  a different approach using
 the   assumption that the  supersymmetric
  \renyi  entropy   for (1,0)  superconformal 6d theory   should be 
   is a  cubic polynomial in  inverse of its argument.   %$q^{-1}$. %   inverse of its argument. 
  %AT28 ????

  %  for unitary 
   %theories.  
 %    based on  supersymmetric  \renyi  entropy 
 %   as an additional input  a conjecture  about a particular   polynomial 
 %  dependence of the supersymmetric  \renyi  entropy on inverse of its argument. 
   %solution 
%also belong to the above 1-parameter family but for  a different value of $\xi=  -\frac{8}{9}$.

In this paper  we   will  settle the    question  about    the   right   value of $\xi$  
in   our  original approach of \cite{Beccaria:2015ypa}
by using   an additional   information about 
the anomalies of  the  free $(1,0)$  vector supermultiplet. We   will  confirm   that  
 the value \rf{188} suggested   in  \ci{Yankielowicz:2017xkf}   %\cite{Beccaria:2015ypa}
is  indeed the correct one. 

 This  multiplet  is  the  higher-derivative (non-unitary)  superconformal 6d  $(1,0)$  
vector multiplet  $V^{(1,0)}$ that  contains
 the 4-derivative  gauge vector  $V^{(4)}_\mu  $  (with 
action $\sim \int F_{\m\n} \del^2  F^{\m\n} $),  the 3-derivative  MW  spinors   $\psi^{(3)}  $  
and the 2-derivative scalars $\vp$  \cite{Ivanov:2005qf,Beccaria:2015uta,Beccaria:2015ypa} (cf. \rf{1.6}) 
\be\la{1.10}
V^{(1,0)} = 3\varphi+2\psi^{(3), +}+V^{(4)} \ . 
\ee
The anomaly polynomial for this multiplet has coefficients\footnote{The  corresponding a-anomaly is $\ \aa(V^{(1,0)} )=-\frac{251}{210}$
\ci{Beccaria:2015uta,Beccaria:2015ypa},  in agreement with (\ref{1.5}).}
\be
%V^{(1,0)}:\qquad
 (\alpha, \beta, \gamma, \delta) = (-1, -\tfrac{1}{2}, -\tfrac{7}{240}, \tfrac{1}{60}),
\ee
so that  using (\ref{1.7}),\rf{1.2}, we  should thus expect to find 
\be\la{1.11}
%V^{(1,0)}:\qquad
\cc_{1} = \tfrac{40}{189}-\tfrac{3}{7}\,\xi\ , \qquad\ \ \ 
\cc_{2} = \tfrac{551}{1890}+\tfrac{3}{28}\,\xi\ , \qquad\ \ \ 
\cc_{3} = \tfrac{13}{210}+\tfrac{3}{28}\,\xi\ .
\ee
The  direct computation of Weyl anomalies $\cc_i$  for   $V^{(1,0)}$ %for the  vector and spinors is 
on   a general curved background  is  challenging 
 as it  requires the knowledge of  6d  Seeley-DeWitt    coefficients  for the corresponding 
 higher derivative   vector and spinor  operators.
 %v3
 In the two special cases -- of  a sphere  and a  Ricci flat   space -- 
   that were   discussed  
 in \ci{Beccaria:2015uta,Beccaria:2015ypa}  the higher derivative  operators factorize 
 and the anomalies  can be readily computed  using the  expressions for the 
 Seeley-DeWitt   coefficients   of  2nd order Laplacians. This fixes  3   %2 (or 3, if one assumes \rf{1.2}) 
  out of 4 coefficients in \rf{1.1}  and thus   does not allow to determine $\xi$. %\foot{
  %v3
  In fact, the  higher derivative scalar, vector and (squared) spinor operators  discussed  below factorize 
  also on a general Einstein  space $R_{\m\n} =  {1\ov d} g_{\m\n} R$ (on which the curvature invariants  in \rf{1.1} remain independent) 
  and thus   their  6d anomalies   may  be computed using  the  6d   Seeley coefficient
   of  2nd order  Laplacians as  was done  in \ci{Bastianelli:2000hi} for  the standard  scalar, spinor
    and 2-form fields.\foot{We thank 
   D. Diaz  for this remark.}     We will   use this observation  below in Section 6. 
     %We will not follow  this route here.

   To fix   the value of $\xi$  it is sufficient,  according to \rf{1.11},   to compute  
   just  $\cc_3$    which itself is determined by the coefficient $C_T$ 
appearing in the 2-point function of stress tensor in flat background.
In fact, as $C_T$   for the scalar  and  the 4-derivative vector  is    already known  \cite{Osborn:2016bev,Giombi:2016fct},  it  remains  only  to  compute    it  for  the 3-derivative 
spinor field.\footnote{We thank Ying-Hsuan Lin  and Chi-Ming Chang for this suggestion.}

 In more detail,   the  6d  Weyl anomaly coefficient $\cc_3$ in  \rf{1.1}    is   given by\foot{In  4 dimensions  the coefficient $\cc$   of the Weyl-squared term in the trace anomaly is 
 given by $\cc = { 1\ov 160} C_{T, 4}$.\la{foot8}}
%in six dimensions, we have the relation 
\be
\la{1.13}  \cc_3 =  \tfrac{ 5}{3\times  7!} 
C_{T, 6}\ ,   % = \frac{3}{5}\,7!\,c_{3},
\ee 
where $C_{T, d}$  is the   coefficient in %enters the universal form of the  
the  2-point function of  stress tensor in  a $d$-dimensional CFT 
\begin{align} \la{1.14}
&\qquad \qquad \qquad  \langle T^{\mu\nu}(x)\,T^{\rho\sigma}(0)\rangle = {}{}\,\frac{C_{T, d}}{[\V_{S^d}]^{2}\, (x^{2})^{d}}
\,\mc I^{\mu\nu \rho\sigma}(x)\ , \\ %Here $\mc I$ is the inversion  operator % for symmetric traceless 2-tensor
& \V_{S^{d}} = \frac{2\,\pi^{d\ov 2}}{\Gamma({d\ov 2})} \ , \qquad 
 \te 
\mc I^{\mu\nu \rho\sigma} = \frac{1}{2}(I^{\mu\sigma}\,I^{\nu\rho}+I^{\mu\rho}\,I^{\nu\sigma})
-\frac{1}{d}\,\eta^{\mu\nu}\,\eta^{\sigma\rho}, \quad I^{\mu\nu} = \eta^{\mu\nu}-\frac{2}{x^{2}}\,
x^{\mu}\,x^{\nu}\ . \no %\la{1.15}
\end{align}
The coefficient $C_{T,d}$ is known for several unitary and non-unitary conformal theories
\cite{Osborn:1993cr,Petkou:1994ad,Petkou:1995vu,Buchel:2009sk,Diab:2016spb,
Giombi:2016fct,Osborn:2016bev}.
In particular, for  the standard  real conformal scalar and    spin $1/2$ fermion one has 
\be
\la{1.16}
C_{T,d}(\varphi) = \tfrac{d}{d-1}, \qquad \qquad  C_{T,d}(\psi) = \tfrac{1}{2} n_{f}\,{d}\ ,
\ee
where $n_{f}$ is the (complex) dimension of the  spinor space 
 ($n_f=2^{{d\ov 2}-1} $ for Majorana  and   $2^{{d\ov 2}-2} $ for MW case). For example, 
 for a 4d Majorana  fermion  $ C_{T,4}(\psi) = 4$, while 
for  a 6d MW  fermion   $C_{T,6}(\psi) = 6$.\foot{Let us add also that  for  an   (anti) self-dual 6d tensor 
one finds $C_{T,6} (T^+ ) =54$.
The corresponding values of $\cc_3$  in \rf{1.13}   are 
$\cc_3(\varphi) =  \tfrac{1}{2520}, \ \ 
\cc_3(\psi) = \tfrac{1}{504} , \ \ 
\cc_3 (T^+)  = \tfrac{1}{56}.$}

 Using the scalar value in   \rf{1.16} and the known   value of  $C_T$   for the  4-derivative gauge   vector $V^{(4)}$  \cite{Osborn:2016bev,Giombi:2016fct}  %and  that for  a MW   6d spinor   $n_f=2$
 % we have    
  \be  \la{115}  C_{T,6}(V^{(4)}) = -90 \ ,  
  %qquad C_{T,6}(\psi^{(3)+}) = \kappa\,C_{T,6}(\psi^{+}) =6 \k \ ,  \la{116}
  \ee 
  %where $\k$  is a constant to be determined. 
% Then 
 we find  that  $C_T$  for the 4-derivative vector multiplet \rf{1.10}  is   given by 
\be
\la{1.18}
C_{T,6}(V^{(1,0)}) = 3\times\tfrac{6}{5}+2\times  C_{T,6}(\psi^{(3)})   -90 \ . %= -\tfrac{432}{5}+12\,\kappa  \ . 
\ee
Comparing this to \rf{1.11}, \rf{1.13} we conclude 
that  the   two   suggested   values of $\xi$ in \rf{1.8}   and \rf{188} correspond to 
\begin{align} \la{1180}
&\xibt   =- \tfrac{31}{27} \ \ \ \  \ \ \ \to \ \ \ \ \ \ \ 
 C_{T,6}(\psi^{(3)})   = -\tfrac{246}{5}  \ , \\
&\xiyz  =- \tfrac{8}{9} \ \ \ \  \ \ \ \ \to \ \ \ \ \ \ \ 
 C_{T,6}(\psi^{(3)})   = -\tfrac{36}{5}  \  . \la{118} \end{align}
As we shall   find below,  it  is   the second   value   \rf{118}
that  is  the correct  result for the 
$C_T$  of  the 3-derivative  6d MW  fermion.

%by  the direct computation of 
%$C_T$ for the 3-derivative  6d MW spinor field
% that we  will  find below , thus confirming the relations  \rf{1.9}.
 
 %\foot{It is interesting to note that  while  the 
%values of $C_T$ for the  standard unitary  fields are positive, its  values 
%for 4-derivative non-unitary fields  appear to be  always negative. The values  of $C_T$ 
%for 6-derivative scalar  are again positive (see \rf{a1}).}

   To find   $C_T$  for a  free conformal field  one  may follow the standard route of  first determining 
the   explicit  form of the  stress tensor $T_{\mu\nu}$    as a conformal primary or 
%which can be  
 obtaining it  from the  metric 
variation of  a Weyl-invariant  action in curved  background
and then using \rf{1.14}.\foot{Some    simplification is that in  computing   $\langle T T\rangle $ in  \rf{1.14}  one  can drop total derivative (descendant)   parts   in one of the two  $T$-factors \cite{Osborn:2016bev}.} 
An alternative  approach  that we shall follow   below 
  is to exploit the relation between $C_{T}$
and  the \renyi entropy \cite{Perlmutter:2013gua}.
 As   we shall  demonstrate, this second approach 
turns out to be more  efficient  in the case of the  higher-derivative  conformal fields.
% like  the 3-derivative spinor   field. 
%We will also  
%   illustrate  its efficiency   on the examples 
%of  higher derivative scalar and  vector conformal fields. 

\def \F  {{\cal F}}

Given  a CFT in flat  even-dimensional  
space   one has  the following relation between the
first derivative   of the  \renyi
entropy $\SS_{q}$  (which is a function of $q$ defined in the next section)   at $q=1$ 
 and  the coefficient $C_{T,d}$ in  \rf{1.14} \cite{Perlmutter:2013gua}
% ($\SS'_1 \equiv \SS'_q\big|_{q=1}$)
\be
\la{2.1}
\SS'_1= -\VV_{\mathbb H^{d-1}}\,\frac{\pi^{\dd+1}\,\Gamma(\dd)\,(d-1)}{(d+1)!\ [\V_{S^d}]^{2}}\, \,C_{T,d}  \ . %, \qquad S_{d} = \frac{2\,\pi^{d/2}}{\Gamma(d/2)}.
\ee
Here  $\V_{S^d}$ is the volume of the  sphere  as in   \rf{1.14}
and $\VV_{\mathbb H^{d-1}}$ 
is the  finite coefficient  in the   regularized volume of  the odd-dimensional unit-radius 
 hyperbolic space $\mathbb H^{d-1}$  
 ($\LI$ is an IR cutoff)   %is, see (3.1) of \cite{Perlmutter:2013gua}, 
\be
\la{2.2}
\V_{\H^{d-1}} \equiv \VV_{\mathbb H^{d-1}}\,\log{ \LI}\ , \qquad \qquad 
\VV_{\mathbb H^{d-1}} = (-1)^{\dd-1}\, \frac{2\pi^{\dd-1}}{\Gamma(\dd)}\ . 
\ee
In particular, 
\be \la{488}
\VV_{\mathbb H^{3}} = -2\pi, \ \qquad
\VV_{\mathbb H^{5}} = \pi^{2}, \ \qquad 
\VV_{\mathbb H^{7}} = -\tfrac{\pi^{3}}{3}\ , \ \ \ \   ...\ee
%e.g., $\V(\mathbb H^{5}) = \pi^2     $.
 and thus   %This gives %(for even $d$) 
%\be\la{2.3}
%d\in 2\,\mathbb N: \quad 
%C_{T,d} =\frac{2\, (-1)^{\dd}\, (d+1)!}{(d-1)\,[(\frac{d}{2}-1)!]^{2}}\,\SS'_1 \ , 
%\ee
%For instance, in $d=4,6,8$, we have i.e. 
\be
\la{2.4}
\ \quad \ C_{T,4} = 80\,\SS'_1, \qquad
C_{T,6} = -504\,\SS'_1, \qquad
C_{T,8} = 2880\,\SS'_1 \ , \ \  \ \ \   ...
\ee
%%%%%%%%%%%%%%%%%%%%%%%%%%%%%%%%%%%%%%
%For completeness  we will also  treat  scalar and vector   cases, etc.
We shall start in  section 2   with defining the \renyi entropy  in terms of the  free energy $\F_q$ on  $S^1_q \times \H^{d-1}$, i.e. 
the product of a  thermal circle (with length $\beta = 2 \pi q$) 
and  hyperbolic space, thus relating $C_T$ to  the second derivative of the free energy at $q=1$. 
We will  then describe  our method  of computing this  free energy  using heat  kernel representation.

To  illustrate this method   of computing  free energy   and  $C_T$  %for  generic free conformal fields  
  %and  %explore  its application to  the case of higher derivative 
%conformal fields 
in  section  3 we  will  consider  the examples of  the 
4- and 6-derivative conformal scalars  in even  number of dimensions. 
In section 4  we will discuss   the case of the 4-derivative   conformal  gauge vector  in   6d
reproducing  the value  \rf{115}  of its $C_T$ obtained earlier by other methods. 

Section 5 will  contain a similar computation  of  free energy and thus  \renyi entropy and 
$C_T$ for the 3-derivative conformal fermion   in $d=4$  and $d=6$.
%deriving the value  given  in \rf{118}. 
For higher derivative operators  the computation of  $C_T$  
turns out to be  subtle: surprisingly, 
a  naive approach (discussed  in Appendix A) leads to the value  in \rf{1180}   while the correct evaluation  gives  \rf{118}. 

%v3
In section 6 we will provide an independent derivation of the  conformal anomalies of the vector multiplet \rf{1.11}  with $\xi$ given by \rf{188} 
by  directly computing the Seeley-DeWitt    coefficients of the   higher derivative  operators   involved using the fact of  their factorization on a  generic  Einstein background.

In Appendix A we will  supplement the  discussion in 
section 5   by  explaining a different method  of computing the
 free energy on $S^1_q \times \H^{d-1}$.  
 In Appendix B we will  compute $C_T$   for  the non-unitary 2-derivative 
conformal  vector theory   which    has no   gauge invariance in   $d\not=4$. 
%mb
Finally, in Appendix C, we shall present the result for the conformal anomalies for a family of 
vector multiplets generalizing \rf{1.10}  that  shows again  the agreement with the relations \rf{1.2}, \rf{1.7} with  
$\xi$ given by  \rf{188}.

\def \OO   {{\cal O}} 
\def \F {{\mc F}}

%%%%%%%%%%%%%%%%%%%%%%%%%%%%
\section{Free energy   for  conformal fields   on  $S^{1}_{q}\times \mathbb H^{d-1}$ 
and  \renyi entropy
   \la{sec2}}
%%%%%%%%%%%%%%%%%%%%%%%%%%

The \renyi entropy $\SS_{q}$  is a measure of generalized quantum entanglement and 
can be computed from traces of the reduced density matrix raised to a power $q\ge 0$.
For  a  $d$-dimensional CFT, the \renyi entropy  across $S^{d-2}$ may be equivalently 
extracted from the  partition function on $q$-cover of the sphere 
$S^{d}$ or  from the   thermal partition   function 
on $S^{1}_{q}\times \mathbb H^{d-1}$  (see \cite{Casini:2011kv,Klebanov:2011uf,Lewkowycz:2014jia}
and refs. there).\foot{The  metrics  of the two spaces are  related  by a singular  conformal rescaling
\be \no ds^2_{ q\rm -sphere} =  \sin^2 \theta\,   q^2  d \tau^2  + d \theta^2 + \cos^2\theta\ d\Omega^2_{d-2}
   = \sin^2 \theta\,  \big(  q^2  d \tau^2  + d \rho^2  + \sinh^2\rho\,  d\Omega^2_{d-2}\big)
   = \cosh^{-2 }\rho\,  ds^2_{S^1_q \times \H^{d-1}} \ee
   Here $\tau\in (0, 2\pi) $   and   $\sinh\rho=\cot \theta$.
   This transformation    maps  the  %invariant  
   subspace   $S^{d-2}$   to the boundary of 
$\mathbb H^{d-1}$. 
   For $q=1$ the space $S^{1}_{q}\times \mathbb H^{d-1}$   becomes conformal 
 to  regular $S^d$   and thus also to $\mathbb R^d$  
  as
   $ds^2= dz^2 +   z^2  d x_0^2 +  dx_n dx_n = z^2 \big(dx_0^2 + { {dz^2 + dx_n x_n} \ov z^2}
\big)$.   
   \la{foot11}}
Here $\mathbb H^{d-1} $ is real hyperbolic space
(of  curvature radius $r=1$) 
and the length of the  thermal circle $x_0= q \tau$  or the inverse temperature  is $\beta=2\pi q$.

%%%%%%%%%%%%%%%%%%%%%%%%%%%%%%%%
\subsection{General relations}

Here we shall  use the  latter  definition of $\SS_q$ in terms of the partition function or free energy  on 
$S^{1}_{q}\times \mathbb H^{d-1}$  for even $d$. 
 Given    a  free  real  conformal field $\Phi$ % (with possible flavour or spin indices) 
 with the   action 
\be
\la{3.1}
I = \tfrac{1}{2}\,\int %_{S^{1}_q\times \mathbb H^{d-1}}
d^{d}x\,\sqrt{g}\,\Phi\,\OO\,\Phi,
\ee
where $\OO$ is a (possibly higher order)  covariant  differential operator including curvature terms  needed to  ensure  the Weyl invariance of \rf{3.1}  in a general curved  background, 
the corresponding   free energy  on ${S^{1}_q\times \mathbb H^{d-1}}$   is  %and  the \renyi entropy  are 
\begin{align}
\la{32}
&F_{q} = -\log Z_{q} = \tfrac{1}{2}\,\log\det \, \OO \ .   %\equiv \F_q \log \LI \  , \\
%&S_{q} \equiv  \frac{q\,\F_{1}-\F_{q}}{1-q} \ , \la{32}
\end{align}
In  the present  case of   a   homogeneous   space $F_q$  is proportional to its volume, i.e. to 
$ 2\pi q\,  \V_{\H^{d-1}} $  in \rf{2.2}. Extracting the IR divergent factor, we may 
define the  IR finite  "free energy" 
$\F_q$  by 
\be\la{321}   F_q \equiv  \F_q \log \LI \ . \ee
For even $d$  the free   energy   on  ${S^{1}_q\times \mathbb H^{d-1}}$   
 does not contain logarithmic UV   divergences\foot{
% and contains also logarithmic UV divergence  proportional to $\aa$-anomaly coefficient  times the  constant Euler density of $S^{1}_{q}\times \mathbb H^{d-1}$.\foot{
Since 
$S^{1}_{q}$ factor is flat and $\mathbb H^{d-1}$  is conformally flat,   all 
logarithmic   divergent terms  containing the Weyl tensor vanish, while the   Euler density in $d$  dimensions  vanishes 
when evaluated on $\mathbb H^{d-1}$.}
    while the  non-universal power divergent  part of $\F_q$  (which is proportional to  the volume and is  thus    linear   in $q$) 
 should be subtracted  using some regularization  prescription. 
%Thus (dropping power UV divergences and with $\LU$ being UV cutoff in units of curvature radius) 
%\be\la{33}
% F_q \sim\  q\,  \V(\H^{d-1})  \  \aa \, \log \LU  + ... \ , \ee where  dots  stand  for finite  terms vanishing for $q=1$ (when the space becomes conformal  to $\mathbb R^d$). 

%We may   define the  IR finite  "free energy" 
%$\F_q$   obtained from $F_q$ by omitting   the  IR divergent  factor $\log \LI$
%in the volume of  $\mathbb H^{d-1}$     as in \rf{2.2};  then 
 The finite  \renyi entropy
is then given by 
\be \la{3.2}
\SS_{q} \equiv  \frac{q\,\F_{1}-\F_{q}}{1-q}  \ , \ \ \ \qquad  \ \ \ \ \    
\F_q =  q \F_1 +  ( q-1)\, \SS_q  
\ . \ee
%In view of \rf{33}  the \renyi entropy $\SS_q$  is UV finite  and 
%where $\F_1$   is a scheme-dependent constant  and $\SS_q$ term represents the finite
%unambiguous part. 
Note that   under a  linear in $q$   and constant   shift of the free  energy we have 
\be \la{2244}
\F_q \to  \F_q  +  k_1  q  + k_2   \ \ \ \   \to  \ \ \ \ 
\SS_q  \to \SS_q    + k_2 \ . \ee
As all power UV divergent terms   in $\F_q$ are linear in $q$   they 
drop out  of $\SS_q$  which is 
thus UV finite. 
The $q=1$ value of the \renyi   entropy   which  is  the entanglement entropy
\be \SS_{1}= \F'_{1}-\F_1  \la{373}
 \ee
 is   sensitive to  the constant  ($q$-independent) part of $\F_q$. 
%A  UV subtraction   scheme dependent constant  term in $\SS_q$  may be fixed   by demanding 
$\SS_1$   is  
   expected to  be   proportional to the $\aa$-anomaly coefficient of the $d$-dimensional CFT, e.g.,\foot{In 4 dimensions  (cf. \rf{1.1}) $(4 \pi)^2 T^\m_\m = - \aa\,   R^*R^*  + \cc\,  C^{\m\n\l\r} C_{\m\n\l\r} $.}
   \be 
 d=4: \ \   \SS_1 = - 4 \aa \ , \ \ \ \ \qquad\qquad   d=6: \ \  \SS_1 = -96 \aa \ , \la{555}
   \ee
 as that  happens   when $F_d$  is computed on  the $q$-cover of the sphere 
$S^{d}$ 
\ci{Solodukhin:2008dh,Dowker:2010bu,Dowker:2010nq,Solodukhin:2010pk,
Casini:2011kv,Aros:2014xga}.\foot{One  expects that the  log UV  divergent  part of free energy on  $q$-cover of the $S^{d}$  should be  matching  the 
 log IR  part of free energy on $S^1_q \times \mathbb H^{d-1}$,  and that was checked on specific examples, though a general proof of this statement appears to be  missing in the literature.}
However,  the   transformation between   the 
$q$-cover of the $S^{d}$  and  $S^1_q \times \mathbb H^{d-1}$
is a non-trivial   Weyl  rescaling (cf. footnote \ref{foot11})
and  thus  the two free energies may a priori differ by a  Weyl-anomaly term. % i.e.
  % by  a constant. 
%While   this does not happen for scalar and spinor fields,   
It was observed that  for  fields   with gauge  invariance   $\SS_{1}$ 
  computed on ${S^{1}_q\times \mathbb H^{d-1}}$   is not  automatically proportional to the 
 Weyl anomaly  a-coefficient  (see 
  \ci{Donnelly:2014fua,Huang:2014pfa}  for  4d vectors  and  \ci{Nian:2015xky}  for 6d antisymmetric tensors), 
  but one can  achieve this  by shifting $\F_q$ by a constant
  (that  may be interpreted as  an edge mode contribution). 
 % Shifting $\FF_q$ by a constant one can always   make 
 
   The $C_{T}$   coefficient 
 which   is proportional to the first derivative of the \renyi entropy \rf{2.1}
  may   be expressed in terms of the    second  derivative 
of the free  energy  $\F_q$     and thus is not sensitive to the shifts  in \rf{2244}.
Explicitly,  
\be
\la{3.3}
C_{T,d} = \tfrac{(-1)^{\dd}\, (d+1)!}{(d-1)\,[(\frac{d}{2}-1)!]^{2}}\, \mc F_{1}''\  , \   \ \ \ \ \ \ \  \ \qquad \ \ \ 
\SS'_{1} = \tfrac{1}{2}\,\mc F_{1}'' \ .
\ee
In particular  (see \rf{2.4},\rf{1.13}   and \ref{foot8})
\be 
d=4: \ \ \ C_{T,4} = 160\,  \cc= 40\,  \F_{1}'' \ , \ \ \ \ \qquad \ \ \ \ \ \ 
 d=6: \ \ \ C_{T,6} = 3024\,  \cc_3=  - 252\, \F_{1}''  \ . \la{3346}
\ee
%The UV divergent part of  $F_q$ being local does  not contribute to \rf{3.3}.
%In fact, the  $q\to 1$  limit of   $\SS_q$     which  is the entanglement entropy  $S_{EE}$ 
%is proportional to the $\aa$-anomaly coefficient  \ci{Myers:2010tj} (which is the coefficient  of the 
% UV divergent part of  free energy on $S^d$). 
Thus to compute $C_T$ we need to find 
the free  energy  $\F_q$  on $S^{1}_q\times \mathbb H^{d-1}$.
%Thus to find $C_T$ we 

%%%%%%%%%%%%%%%%%%%%%%%%%%%%%%%%%%%%%%%%%%%%%%%%
\subsection{Computational scheme}
%%%%%%%%%%%%%%%%%%%%%%%%

The   covariant kinetic  operator $\OO$   specified to $\Ss^1_q \times  X^{d-1}$
where $X^{d-1}$  is a symmetric space like $S^{d-1}$ or $\H^{d-1}$ 
will be a polynomial  in  derivatives  $\del_0$ 
along the "euclidean time" direction $\Ss^1$  and  the 
  covariant derivatives $D_i\equiv \bm{D}_{i}$  on  $X^{d-1}$, 
  i.e.    symbolically $\OO = P (i\,\partial_{0}, -\bm{D}^{2})$
(with $X^{d-1}$ curvature  factors   translating into  the 
  coefficients  of lower-order terms in $P$). 
In the case of  $X^{d-1}= \H^{d-1}$  the   free energy  $\F_q$ in \rf{32},\rf{321} 
   will  have the following   structure
 %take the typical form 
\be
\la{3.4}
\F_{q} = \ha {\VV_{\mathbb H^{d-1}}} \,\sum_{n}\int %_{\lambda_{0}}^{\infty}
d\mu(\lambda)\,\log  P_\H\big(\frac{n}{q}, \lambda\big), 
\ee
where  $\frac{n}{q}$ is the eigenvalue of $i\d_0$    and 
$d\mu(\lambda)$ is the spectral measure for the continuous eigenvalue $\lambda$ of  
the  spatial  operator 
$-\bm{D}^{2}  + ...$ (a particular definition of $\l$  will    depend
on  a  type of the field $\Phi$ in \rf{3.1}, see below). 
%  Its specific form depends in particular on the spin of $\Phi$.  
The summation  index   $n$  takes values  in $\mathbb Z$ for bosons and in $\mathbb Z+\frac{1}{2}$
for fermions.\foot{The  antiperiodicity of fermions  in  "thermal" circle 
is related to  the   original definition of partition function on $q$-cover of $S^d$.}

It turns out that  for conformal fields the kinetic operators  $\OO$   restricted to $\Ss^1_q \times  X^{d-1}$, i.e. 
 $ P(i\,\partial_{0}, -\bm{D}^{2})$, 
have  special factorized  structure,   i.e. are given by a product of simple 
 two-derivative factors.\foot{This applies to bosonic operators and squared fermionic operators.} 
  A particular  reason for  this  can be understood by observing that 
the  operators  on  $\Ss^1 \times \mathbb H^{d-1}$   and  $\Ss^1 \times  S^{d-1}$
are  formally related by an analytic continuation  changing the sign of the curvature. 
The thermal   partition  function on $\Ss^1 \times  S^{d-1}$   is expressed in terms of characters 
of conformal group  and this in turn is  related to   factorization of the  (higher-derivative) kinetic operator 
discussed in  detail in \cite{Beccaria:2014jxa}. 
In the case of $\Ss^1_q  \times  S^{d-1}$  
 we get 
\be
\la{3.5}
%S^{1}\times S^{d-1}:\qquad \mc 
F_{q} =\ha  \V_{S^{d-1}}\,
\sum_{n}\sum_m  %_{m=m_{0}}^{\infty}\,
\mu(m)\,\log P_{S}\big(\frac{n}{q}, m\big), 
\ee
where the sum over $m$ is over the discrete spectrum of $-\bm{D}^{2}+...$ on $S^{d-1}$ and 
$\mu(m)$ is  the multiplicity factor  of the  eigenvalue  with label $m$. 
  %As one   can see in many examples,  
 The  higher-derivative  Weyl-covariant  operators   $\OO=  D^{2p } + ...$ turn out to factorize
 \cite{Beccaria:2014jxa} into simple factors  so that  the corresponding eigenvalues on
    $S^1_q\times S^{d-1}$ are 
\be
\la{3.6}
%S^{1}\times S^{d-1}:\qquad
 P_{S} = \prod_{k=1}^{p} %\prod_{m=m_{0,k}}^{\infty}
\big[\frac{n^{2}}{q^{2}}+ \frac{1}{r^2}(m+\ell_{k})^{2}\big] \ , 
\ee
where $r$ is the radius of $S^{d-1}$.
%\foot{Eqs. \rf{3.5},\rf{3.6} 
%are given in schematic form -- the lower  value of $m_0$  in \rf{3.5}    may actually be 
%different for each $k$-factor.} 
%where $\OO_{S} = (\partial^{2})^{N}+\cdots$.
In this case, % the standard manipulations show
the standard  free energy  $F_q$ in \rf{32}   is expressed  in terms of 
 the  single-particle partition function $\mc Z(x)$ 
that  has a  simple structure 
\be
\la{3.7} F_{q} = \sum_{n=1}^{\infty}\frac{1}{n}\,\mc Z(x^{n})\ , \qquad \quad 
\mc Z(x) = \sum_{k=1}^{p}\sum_m %{m=m_{0,k}}^{\infty}
\mu(m)\,x^{m+\ell_{k}} \ , \ \ \
 \qquad \qquad x\equiv e^{-2\pi q} \ . 
\ee
Here  $m+\ell_{k}$   correspond to  the single-particle energies 
or integer dimensions of   conformal operators 
 in $\mathbb R^d$   built out of   $\Phi$  and its derivatives. 
%of a set of $N$ 2-derivative fields  related to $N$ $\Phi$. 

The factorization of the higher-derivative  Weyl-covariant kinetic  operator $\OO$ on $\Ss^1 \times \mathbb H^{d-1}$  is thus  intimately 
related  to   its factorization on $\Ss^1 \times  S^{d-1}$  which in turn is related to 
integrality of dimensions of  the  CFT operators in $\mathbb R^d$.\foot{Similar  factorization is found   also for $\OO$  defined on  $S^d$  or $\H^d$.}

\def \ed {\end{document}}
\def \sql {{\sqrt \lambda}}
\def \bdd {{\bm{ \Delta}}}

One may also  consider the analytic continuation  between $S^{d-1}$ and $\H^{d-1}$ 
 at the level of the   spectrum (see \cite{Camporesi:1994ga,Gopakumar:2011qs}
 and  Appendix  C of 
\cite{Beccaria:2014xda}). 
For example,  for   a    2nd order Laplacian  acting 
 on 
symmetric traceless rank $s$  tensors  on a homogeneous  space    one has  the following spectrum 
on $S^{d-1}$  with  radius $r$ 
%that has the following generic form  
\be \la{x}
-  \bm{D}^{2}_{S^{d-1}} %+\alpha\,\overline R  %frac{R}{(d-1)(d-2)}
\,\vp_s = \omega_m \,\vp_s \  , \qquad 
\omega_m=  \tfrac{1}{r^{2}}\,\big[(m+  \tfrac{d-2}{2} )^{2} -   ( \tfrac{d-2}{2} )^2 -s  \big], 
\qquad  m=s, s+1, \dots \ . %%  \\
%\rho &= \tfrac{d-2}{2}, \qquad b^{2} = \rho^{2}+s-\alpha.\no 
\ee
 The eigenvalues  $\omega_{\lambda} $  
of the same operator on $\mathbb H^{d-1}$  with  curvature radius $r$ are  obtained by 
replacing 
\be\la{x1}  m\to i\,\sqrt\lambda-   \tfrac{d-2}{2} \ , \qquad r\to i\,r \ , \qquad \ \omega_m \to 
\omega_{\lambda} = \tfrac{1}{r^{2}}\big[\lambda+  ( \tfrac{d-2}{2} )^2 +s    \big] \ .
\ee
Here  $0 \leq \l <\infty$  is the eigenvalue of the following operator on $\mathbb H^{d-1}$
(here and in what follows we set  the radius of $\H^{d-1} $ to be $r=1$) \cite{Camporesi:1994ga}  
\be \la{x2}\bdd_s \,\vp_s = \l \,\vp_s \ , \ \ \ \ \ \ \  \ \ \ \qquad 
\bdd_s= 
-  \bm{D}^{2}_{\H^{d-1}} -   ( \tfrac{d-2}{2} )^2 - s   %+\alpha\,\overline R  %frac{R}{(d-1)(d-2)}
%\,\vp_s = \l \,\vp_s 
\ . 
\ee
The analytical continuation \rf{x1} then translates  the factorization (\ref{3.6}) into the one on 
 $S^{1}\times \mathbb H^{d-1}$. 
 
 \def \OK {{\overline K}}
 
In addition, we need  to replace the sum 
 $\sum_{m}\mu(m) $ in \rf{3.5}  by $  \int d\mu(\lambda)$ in \rf{3.4}  with a definite correspondence between the 
discrete multiplicity on $S^{d-1}$  and the spectral measure on $\H^{d-1}$.  The latter 
 is  the  Plancherel measure for  the  transverse traceless symmetric rank $s$  field  
 on $\H^{d-1}$   corresponding to  the spectrum  \rf{x2} 
 \cite{Camporesi:1994ga}
 \be\la{5.9}
d\mu_{s, d-1} = \frac{(2s+d-4)(s+d-5)!}{(d-4)!\ s!}\,\frac{\l
+(s+\frac{d-4}{2})^{2}
}{2^{d-2}\,\pi^{d-1\ov 2}\,\Gamma(\frac{d-1}{2})}\,\left|
\frac{\Gamma(i\,\sql+\frac{d-4}{2})}{\Gamma(i\,\sql)}\right|^{2}\,d\sql \ .
\ee
% In what follows we set the radius of $\H^{d-1}$ to 1. 
Having   $\OO$  factorized into a product of  second-derivative factors,  
the polynomial $P_\H$ in \rf{3.4}  may   be 
 written in the   product form  which is the counterpart of \rf{3.6}, 
\be
\la{x3}
%S^{1}\times S^{d-1}:\qquad
 P_{\H} = \prod_{k=1}^{p} %\prod_{m=m_{0,k}}^{\infty}
\big[\frac{n^{2}}{q^{2}}+ (\sql + i  \a_{k})^{2}\big] \ , 
\ee
where $\a_k$ are real  constants (appearing in $\pm$ conjugate pairs so that $P_\H$ is real).
Then $\log P_\H$ in \rf{3.4}   becomes the sum of $p$ terms. Using the 
proper-time representation  separately for each log term  in the sum 
we  then get  (in bosonic case)
\begin{align} 
\la{x4}
&\F_{q} =  %\sum_{k=1}^p \F_q^{(k)} \ , \ \ \ \ \qquad \quad  \F_q^{(k)} =
 -  \ha \, {\VV_{\mathbb H^{d-1}}} \, 
  \int ^\infty_0 {dt \ov t}  \, K_{S^1}(t) \, \overline  K_{\H^{d-1}}(t) 
%\int^\infty_0 d\mu(\lambda)\,    e^{- t  (\sql + i  \a_{k})^{2} }
  \ , \qquad \qquad K_{S^1}(t) = \sum_{n\in \mathbb Z } e^{- t {n^2\ov q^2}} \ ,\\
& \la{x5}
\overline  K_{\H^{d-1}}(t) = \sum_{k=1}^p \OK_{\H^{d-1}}(t;\a_k )\ , \ \ \ \ \qquad 
\OK_{\H^{d-1}}(t;\a_k )=
 \int^\infty_0 d\mu(\lambda)\,   e^{- t  (\sql + i  \a_k)^{2} } \ . 
\end{align}
Here $K_{S^1}$ is the   trace of the 
heat kernel  of $-\del_0^2$ on $S^1$  while $\overline  K_{\H^{d-1}}(t;\a)$
may be interpreted as  the   heat kernel  corresponding to the operator $\big(\sqrt{\bdd_s  } + i \a\big)^2$ on $\H^{d-1}$  (cf. \rf{x2}).
Using the Poisson  resummation\foot{\la{foot17}
In general, 
  \be \no \sum_{n\in \mathbb Z } e^{- t {(n+ \g)^2\ov q^2}} 
=  \frac{2\,\pi\,q}{(4\,\pi\,t)^{1/2}}\sum_{n\in\mathbb Z} \, e^{ 
-\frac{\pi^{2}\,n^{2}\,q^{2}}{t}  + 2 \pi i  \g n }\ \ .
\ee} 
  we may represent $K_{S^1}(t) $ as 
\be
\la{x6}
K_{S^{1}}(t) =  \frac{2\,\pi\,q}{(4\,\pi\,t)^{1/2}}\sum_{n\in\mathbb Z} \, e^{ 
-\frac{\pi^{2}\,n^{2}\,q^{2}}{t}} \ . 
\ee
Similarly, in the   fermion (antiperiodic) case one finds 
\be
\la{x7}
K^{f}_{S^{1}}(t) = \sum_{n\in \mathbb Z+{1\ov 2}  } e^{- t {n^2\ov q^2}}  =  \frac{2\,\pi\,q}{(4\,\pi\,t)^{1/2}}\sum_{n\in\mathbb Z}  (-1)^n  \, e^{ 
-\frac{\pi^{2}\,n^{2}\,q^{2}}{t}} \ . 
\ee
Assuming $t>0$ the integral over $\l$ in  \rf{x5}   is convergent, i.e. the  relevant real part of 
 $\overline  K_{\H^{d-1}}(t;\a)$  is  proportional to  a  finite polynomial in $t$, i.e.
 \begin{align}
 \la{x8}
&\OK_{\H^{d-1}} (t; \a)  + \OK_{\H^{d-1}} (t; - \a)   = \int_{0}^{\infty}d\mu (\lambda)\,
\big[e^{-t\,(\sqrt\lambda+i\,\a)^{2}}+e^{-t\,(\sqrt\lambda- i\,\a)^{2}}\big] \notag \\
& \ \ \ = 2\,\int_{0}^{\infty}d\mu(\lambda)\,
e^{-t\,(\lambda-\a^{2})}\,\cos(2\,\a\,t\,\sqrt\lambda)= \frac{1}{(4\,\pi\,t)^{d-1\ov 2}}
\sum_{j\ge 0} \nu_{j}\,t^{j},
\end{align}
where $\n_j$  are   numerical  constants  depending on  $\a$, dimension $d$  and  spin of the field. 
The integral over $t$ in \rf{x4} is then  power-divergent   at $t=0$ for $n=0$  term in \rf{x6} or \rf{x7}. 
Subtracting these  power divergences  as a proper-time  regularization  prescription 
corresponds to omitting the $n=0$ term in the sum.  As  a result, we are left  with a finite sum over $n\geq1$ expressing $\F_q$ as a finite polynomial in $q^{-1}$   with  coefficients 
 proportional to the Riemann zeta-function values.\foot{ In the antiperiodic  case one has  \be\no 
\zeta_{2k} = \sum_{n=1}^{\infty}\frac{1}{n^{2k}} \quad\to\quad \sum_{n=1}^{\infty}\frac{(-1)^{n}}{n^{2k}} = 
(2^{1-2k}-1)\,\zeta_{2k}\ .
\ee}
 
 To summarize, the  computation of  the free energy $\F_q$  will  contain  the following sequence of steps: (i) integration  over the  eigenvalue $\lambda$; (ii) integration over the proper time $t$  with $t\to 0$  power divergences  subtracted; (iii)   performing the remaining  finite sum  over $n \not= 0$. 
 We shall  illustrate this procedure in detail  on several examples  below.
 Having found $\F_q$  one can then compute the \renyi entropy  in \rf{3.2}    and $C_T$ in \rf{3.3}.

%%%%%%%%%%%%%%%%%%%%%%%%%%%%
\section{Scalar fields}

To  illustrate the relation \rf{2.1},\rf{3.3}   %and  %explore  its application to  the case of higher derivative 
%conformal fields 
in this section we  will use it  compute $C_{T}$ for   free  higher-derivative conformal 
 scalar  theories  in even dimension  $d$, reproducing 
%The  case of the standard 2-derivative conformal
%scalar will  be  review, while
%the  application  of \rf{2.1} to higher derivative
% conformal  scalar
 %This     will   give   a  novel   way to reproduce 
 the 
 results   obtained previously by other methods in a novel way. 

\def \DD {\bm{D}}
\def \sql {\sqrt{\lambda}}\def \mm {{\rm m}}

\def \bdd {{\bm{ \Delta}}}
\def \bd  {{\bm{ \Delta}_0}}

\subsection{$\partial^{2}$ scalar}

The standard action for  the conformally coupled scalar is
\be
I = \tfrac{1}{2}\int %_{S^{1}\times \mathbb H^{d-1}} 
d^{d}x\,\sqrt{g}\,\varphi\,\big[
-D^{2}+\tfrac{d-2}{4\,(d-1)}\,R\big]\,\varphi.
\ee
%Since $R (\H^{d-1}) =-(d-1)(d-2)$, 
The  corresponding   free energy  on $S^{1}_q \times \mathbb H^{d-1}$  is  given by 
($R (\H^{d-1}) =-(d-1)(d-2)$) 
\be
\la{4.2}
F_{q} = \tfrac{1}{2}\,\log {\det}\big(- \del_0^2  + \bd\big) \ , \qquad \qquad 
\bd \equiv - \DD^{2}-\tfrac{(d-2)^{2}}{4} \ .
\qquad 
\ee
The spectrum of the operator  $\bd$ (i.e.  the $s=0$  case of \rf{x2})  is $\frac{n^{2}}{q^{2}} + \lambda$   where 
%$-\partial_{0}^{2}\to \frac{n^{2}}{q^{2}}$ with 
$n\in\mathbb Z$   and $\lambda \ge 0 $. 
The spectral measure is given by  the $s=0$  case of \rf{5.9}, 
in particular, in $d=4$ and $d=6$,  % we get 
\begin{align}\la{42x}
\te 
d\mu_{0, 3} =  \tfrac{1}{4\,\pi^{2}}\sqrt{\lambda}
\,d\lambda, \qquad \qquad 
d\mu_{0, 5} =  
\tfrac{1}{24\,\pi^{3}}\sqrt{\lambda}\,(1+\lambda)\,d\lambda \ .% \qquad \qquad
%d\mu_{0, 7} = \tfrac{1}{240\,\pi^{4}}\sqrt\lambda\,(1+\lambda)(4+\lambda)\,d\lambda.
\end{align}
%is the    eigenvalue of  the "spatial" Laplacian  $\bd$, i.e. 
%\be \la{422}  \big[-\bm{D}^{2}-\tfrac{(d-2)^{2}}{4}\big]\,  \varphi_\l =  \lambda\, \varphi_\lambda 
%\ . \ee
In $d=4$  we get from \rf{x5}
\be
\la{4.17}
\OK_{\H^3}(t) = \int_{0}^{\infty}d\lambda\,\tfrac{\sqrt\lambda}{4\,\pi^{2}}\,e^{-t\,\lambda} = 
\tfrac{1}{(4\,\pi\,t)^{3/2}} \ .\ee
Then using \rf{x6},\rf{488} we find 
\be 
\mc F_{q} = \tfrac{1}{4} q \sum^\infty_{n=1} \int_{0}^{\infty}\frac{dt}{t^{3}}\,e^{
-\frac{\pi^{2}\,n^{2}\,q^{2}}{t}} 
= \tfrac{1}{4}q \sum_{n=1}^{\infty}\tfrac{1}{n^{4}\,\pi^{4}\,q^{4}}  = \tfrac{1}{360\,q^{3}},\la{3535}
\ee
where   we omitted the $n=0$ mode     which corresponds to subtracting 
the $\Lambda^4$  UV  divergence ($t=\eps=\Lambda^{-2} \to 0$).
 The  resulting \renyi  entropy  and the Weyl anomaly  coefficients have indeed the standard values (see \rf{3346}) 
%in agreement with  \cite{Klebanov:2011uf}.
\be\la{3667}
 \SS_{q} = -\tfrac{(1+q)(1+q^{2})}{360\,q^{3}}
\ , \ \ \ \ \ \ \ \ \ \ 
{\rm a} = -\tfrac{1}{4}\,\SS_{1} = \tfrac{1}{360}, \qquad \ \ 
C_{T,4} = 160\, \cc= 80\,\SS'_{1} = \tfrac{4}{3} \ . 
\ee
Similarly, in $d=6$ 
\begin{align}
\la{4217}
&\OK_{\H^5}(t) = \int_{0}^{\infty}d\lambda\,\tfrac{\sqrt\lambda(1+\lambda)}{24\,\pi^{3}}
\,e^{-t\,\lambda} =  \tfrac{3+2\,t}{3\,(4\,\pi\,t)^{5/2}}\ , \\
&\mc F_{q} = -\tfrac{1}{96}q \sum_{n=1}^{\infty}
\int_{0}^{\infty}\frac{dt}{t^{4}}\,(3+2t)\,e^{
-\tfrac{\pi^{2}\,n^{2}\,q^{2}}{t}} = -\tfrac{1}{48\,\pi^{6}\,q^{5}}\sum_{n=1}^{\infty}
\tfrac{3+n^{2}\,\pi^{2}\,q^{2}}{n^{6}} = -\tfrac{2+7\,q^{2}}{30240\,q^{5}}\ , \\
&
\SS_{q} = \tfrac{(1+q)(1+3q^{2})(2+3q^{2})}{30240\,q^{5}}, \ \ \ \ \ 
{\rm a} = -\tfrac{1}{96}\,\SS_{1} = -\tfrac{5}{72\times 7!}, 
\qquad C_{T,6} = -504\,\SS'_{1} = \tfrac{6}{5}\ ,\la{399}
\end{align}
where   we again dropped the $n=0$ term in the sum corresponding to subtracting  the  $\Lambda^6$
and $\Lambda^4$ UV divergences. 
The above  values for $C_{T,d}$ are  in agreement  with the general expression in  \rf{1.16}.

\def \ep {\epsilon}

%%%%%%%%%%%%%%%%%%
\subsection{$\partial^{4}$  scalar}
%%%%%%%%%%%%%%%%%%%%%%%%%
%Let us begin with 4d. 
The Weyl-invariant action  for the 4-derivative scalar in 
  curved  4d  space is  given by \ci{Fradkin:1981jc}  %, see B.22 of \cite{Beccaria:2014xda}, 
\be\la{4.12}
I= \ha \int  d^4x \sqrt g \big[ D^{2}\varphi \,D^{2}\varphi -2\,\big(R^{\mu\nu}
-\tfrac{1}{3}R\,g^{\mu\nu}\big)\,D_{\mu}\varphi 
D_{\nu}\varphi \big].
\ee
The  generalization of  the $D^4$ operator in \rf{4.12}   to
 any   $d>4 $ is the Paneitz operator  \cite{Paneitz:1983}
\begin{align}
& \OO^{(4)} = D^{4}  +    \tfrac{4}{d-2} \,R^{\mu\nu}D_{\mu}D_{\nu}
 + {k}_{d}\,R\,D^{2}  +\tfrac{d-4}{2}\big(
{n}_{d}R_{\mu\nu}R^{\mu\nu}+{m}_{d} R^{2}\big)  + O(D^2R),\la{pan}\\
& \te {k}_{d} = - \frac{d^2 -4d +8}{2(d-1)(d-2)}, \quad  \quad  
{n}_{d} = -\frac{2}{(d-2)^{2}}, \qquad  {m}_{d} = \frac{d^{3}-4\,d^{2}+16\,d-16}{8(d-1)^{2}(d-2)^{2}}\ . 
\notag \end{align}
We  did not  write explicitly  the $D^2R$ term 
as we will be interested in the   homogeneous  $S^{1}\times X^{d-1}$ background
with
\be\la{4444}
R=-(d-1)(d-2)\ep, \qquad R_{0i}=0, \qquad R_{ij} = \tfrac{1}{d-1}R\,g_{ij} = -(d-2)\ep\,g_{ij},
\ee
%Splitting  the operator $D^2=\del_0^2 + \DD^2$ operator as in  (\ref{4.2}),\rf{x2}
%\be
%\la{4.20}
%D^{2} = \partial_{0}^{2}+\bm{D}^{2}, \qquad \partial_{0}^{2} = -\frac{n^{2}}{q^{2}},
%\qquad \bm{D}^{2}= -\lambda-\frac{(d-2)^{2}}{4},
%\ee
%we find the $d$-independent factorization
where    we introduced the  curvature sign  factor $\ep$   which is  $+1$ for 
$X^{d-1}=\H^{d-1}$ and $-1$ for $X^{d-1}=S^{d-1}$. 
Then \rf{pan}    is found to factorize  in either of  the following two  $d$-independent ways 
\ba
\no 
 \OO^{(4)}&=D^{4}+\ha \ep  ({d^{2}-4d+8}) \,D^{2}-4\ep \,\bm{D}^{2}+\tfrac{1}{16}\ep^2 d^{2}(d-4)^{2} 
\no\\ & 
= \big[  (i \del_0 +\sqrt \ep)^2 + \bd %  -   \DD^2   -\tfrac{(d-2)^{2}}{4} 
 \big]\,
  \big[  ( i\del_0 - \sqrt \ep )^2   +\bd  \big]\la{4.21}\\ & 
= \big[  (i \del_0)^2 +  (\sqrt \bd   + i \sqrt \ep)^2  %  -   \DD^2   -\tfrac{(d-2)^{2}}{4} 
 \big]\,
  \big[  ( i\del_0 )^2   + (\sqrt \bd   - i \sqrt \ep)^2   \big]\, ,\la{429}
%\big[\frac{(n+q)^{2}}{q^{2}}+\lambda\big].
\end{align} 
where  $\DD^2 \equiv  D^iD_i$  and $\bd=  -   \DD^2   - \tfrac{(d-2)^{2}}{4}\ep  $ is  the conformal scalar Laplacian    as in \rf{4.2}. 
This  factorization was  already observed   
 on $\Ss^1 \times S^{d-1}$   where $\ep=-1$ 
 (see eq. (B.22) in \cite{Beccaria:2014xda}  for $d=4$). 
 
The  eigenvalues  of $\OO^{(4)}$   are  thus   naturally expressed  in  terms of the eigenvalue
$\l$ 
of the conformal  scalar Laplacian on $\H^{d-1}$ in \rf{x2} 
\be \la{417}\te 
 \OO^{(4)} \ \to \  \big[ ({ n \ov q} + 1)^2+\lambda\big]\,
\big[ ({ n \ov q} - 1)^2+\lambda\big]=\big[\frac{n^{2}}{q^{2}}+(\sql  + i)^2\big]\,
\big[\frac{n^{2}}{q^{2}}+(\sql -i)^2\big] \ . 
\ee
This is thus  the special case of  \rf{x3}  with $\alpha_k=\pm 1$ 
so that the corresponding free  energy can be   computed  as in \rf{x4}--\rf{x8}. 
Explicitly, we  find    that  in this case  $\OK_{\H^{d-1}}(t) $ is given by \rf{x8}  with $\a=1$ 
so that   for  $d=4$  (cf. \rf{3535},\rf{3667}) 
\begin{align}
\la{17}
&\OK_{\H^3}(t) =2 \int_{0}^{\infty}d\lambda\,\tfrac{\sqrt\lambda }{4\,\pi^{2}}
\,e^{-t\,(\lambda-1)}\,  \cos( 2 t\sql) 
=  \tfrac{2-4\,t}{(4\,\pi\,t)^{3/2}}\ , \\
&\mc F_{q} = \tfrac{1-30\,q^{2}}{180\,q^{3}}, \qquad\qquad 
\SS_{q} = \tfrac{(1+q)(-1+29\,q^{2})}{180\,q^{3}},\la{71}\\
&{\rm a} = -\tfrac{1}{4}\,\SS_{1} = -\tfrac{7}{90}\ , \qquad 
\qquad C_{T,4} = 160 \, \cc= 80\,\SS'_{1} = -\tfrac{32}{3} \ .\la{711}
\end{align}
These values of the Weyl anomaly coefficients $\aa$ and $\cc$   for the 4-derivative scalar 
agree with the  result of the direct  computation  in %from the Seeley coefficient   in 
 \ci{Fradkin:1981jc,Fradkin:1985am}. 

In $d=6$ get   (cf. \rf{4217}--\rf{399}) 
\begin{align}
\la{y1}
&\OK_{\H^5}(t) = 2\int_{0}^{\infty}d\lambda\,\tfrac{\sqrt\lambda(1+\lambda)}{24\,\pi^{3}}
\,e^{-t\,(\lambda-1)} \,  \cos( 2 t\sql) =  \tfrac{ 2(1   -{10} \,t) }{3\,(4\,\pi\,t)^{5/2}}\ , \\
&\mc F_{q} = \tfrac{-2+35\,q^{2}}{15120\,q^{5}},\qquad \qquad
\SS_{q} = \tfrac{(1+q)(2+37\,q^{2}+37\,q^{4})}{15120\,q^{5}}, \la{y2}\\
& {\aa} = -\tfrac{1}{96}\,\SS_{1} = \tfrac{4}{9\times 7!}\ , 
\qquad \qquad C_{T,6} =3024\, \cc_3= -504\,\SS'_{1} = 
-6 \ .\la{y3}
\end{align} 
The  value of $\aa$  in \rf{y3}  agrees with the one found in  \ci{Beccaria:2015uta}
(see Table 1  there). 

The  above  values of $C_T$  in \rf{711} and \rf{y3}  are in agreement with the general   expression 
 for   the 4-derivative  conformal scalar  in  dimension $d$  found in 
\cite{Guerrieri:2016whh,Osborn:2016bev}
\be
C_{T, d}(\varphi^{(4)}) = -\tfrac{2\,d\,(d+4)}{(d-1)\,(d-2)}. \la{444}
\ee

%%%%%%%%%%%%%%%%%%
\subsection{$\partial^{6}$  scalar}
%%%%%%%%%%%%%%%%%%%%%%%%%
\def \en {\end{align}}

The general expression for the Weyl-covariant  6-derivative  scalar   operator   in curved 
background  can be found, e.g., in 
  \cite{Osborn:2015rna}.  Ignoring terms  with derivatives  of the curvature 
and specifying to $d=6$   it  can be written as  
\begin{align}
\la{A.2}
\OO^{(6)} =   -D^{6} & -  (16\,P^{\mu\nu} - 6 g^{\m\n} P) D_{\mu}D_{\nu}D^{2}\no \\
 &+8\,( 4P^{\mu\nu}P - g^{\m\n}  P_{\k\r} P^{\k\r} )\,D_{\mu}D_{\nu} +8\,(P_{\m\n} P^{\m\n} -P^{2})\,D^{2}\ , 
\end{align}
where the Schouten tensor  $P_{\m\n}$ and  its trace $P$  are in  general defined  as 
\be 
P_{\m\n} = \tfrac{1}{d-2}\,\big(R_{\mu\nu}-\tfrac{1}{2(d-1)} R\big) \ , \qquad \qquad 
P= P^\m_\m = \tfrac{1}{2(d-1)} R \ . \la{999}
\ee
Using the properties \rf{4444} of the curvature  of $\Ss^1 \times   \H^5$   
%%%%%%%%%%%%%%%%%%
%%%%%%%%%%%%%%%%%%%%%%%%
we find 
\begin{align}\la{a33}
\OO^{(6)} %&= -D^{6}+16\,\bm{D}^{2}\,D^{2}-8\,D^{4}-12\,D^{4}-12\,D^{2}
%+64\,\bm{D}^{2}-32\,D^{2}-20\,D^{2}\notag \\
= -D^{6}+16\,\bm{D}^{2}\,D^{2}-20\,D^{4}+64\,\bm{D}^{2}-64\,D^{2}.
\end{align}
Like  the 4-derivative  scalar operator \rf{4.21},\rf{429} (where $\ep=1$ for $\H^{d-1}$) ) 
  this  operator  may be  factorized
in the  two possible ways 
\ba
\OO^{(6)}
&= \big[  (i \del_0)^2 + \bm{\Delta}_{0}   %  -   \DD^2   -\tfrac{(d-2)^{2}}{4} 
 \big]\,   \big[  (i \del_0 +2)^2 + \bm{\Delta}_{0}   %  -   \DD^2   -\tfrac{(d-2)^{2}}{4} 
 \big]\,
  \big[  ( i\del_0 - 2)^2   +   \bm{\Delta}_{0}  \big]\, \la{f1}\\
  & \la{a3}
  = \big[  (i \del_0)^2 + \bm{\Delta}_{0}   %  -   \DD^2   -\tfrac{(d-2)^{2}}{4} 
 \big]\,   \big[  (i \del_0)^2 + (\sqrt{\bm{\Delta}_{0} } + 2i)^2   %  -   \DD^2   -\tfrac{(d-2)^{2}}{4} 
 \big]\,
  \big[  ( i\del_0 )^2   +   (\sqrt{\bm{\Delta}_{0} } - 2i)^2 \big]\, \ , 
\end{align}
so that  the corresponding eigenvalues  are given by \rf{x3}   with $\a_1=0,\  \a_2=2,\  \a_3=-2$. 
We thus   get  a combination of the  standard 2-derivative scalar and a conjugate pair of operators 
 with the  shift  parameter $\a=2$. The  heat kernel  for  the latter   is given by \rf{x8}
 %  corresponding 
 and as a  result we find  in $d=6 $  (cf. \rf{y1}--\rf{y3})
 \begin{align}
\la{y11}
&\OK_{\H^5}(t) = \int_{0}^{\infty}d\lambda\,\tfrac{\sqrt\lambda(1+\lambda)}{24\,\pi^{3}}
\,e^{-t\,\lambda} \big[ 1  +   2\, e^{4t}  \cos( 4 t\sql) \big] 
=  \tfrac{ 3 - 30 t + 32 t^2 }{\,(4\,\pi\,t)^{5/2}}\ , \\
&\F_q=\tfrac{2-105\,q^{2}+1680\,q^{4}}{10080\,q^{5}},\quad\qquad 
\SS_{q} = \tfrac{(1+q)(2-103\,q^{2}+1577\,q^{4})}{10080\,q^{5}},\la{y12}\\
&{\rm a} = -\tfrac{1}{96}\,\SS_{1}
 =  -\tfrac{123}{8\times 7!}\ ,\qquad  \qquad C_{T,6} = -504\,\SS'_{1} = 
54\ .\la{yy122} 
\end{align}
The  value of $\aa$-coefficient agrees  with the one following  from the  partition function of 6-order GJMS operator 
on $S^6$ \ci{Beccaria:2015vaa}   while   the value of $C_{T,6}$ 
agrees  with the $d=6$ case   of the  general expression  for $\del^6$    conformal scalar  in 
%For a $\Box^{3}$ scalar we first consider $d=6$ because this is critical order in $d=4$ and, for instance,
%the prediction from 
\cite{Osborn:2016bev} 
\be\la{a1}
C_{T,d}(\varphi^{(6)}) = \tfrac{3d(d+4)(d+6)}{(d-1)(d-2)(d-4)}  \ . 
\ee
%is singular for $d=4$ while it is finite and equal to 54 in $d=6$.

%%%%%%%%%%%%%%%%%%%%%%%%%%%%%%
\section{Conformal vector fields } %4-derivative  conformal  6d  gauge vector }
%%%%%%%%%%%%%%%%%%%%%%%%%%%%%%%%%%%%
Conformal   generalization of  the Maxwell theory  to  general dimension $d$  has 
a  higher derivative Lagrangian
$L= F_{\mu\nu} (\partial^{2})^{d-4\ov 2}\,F^{\mu\nu}$. 
In particular, in  6 dimensions    this  gives   a 4-derivative  non-unitary vector gauge 
 theory  that we shall consider below. 
 %As another example  of the application  of the relation\rf{2.1} between $C_T$ and the \renyi 
%entropy  here  we shall  consider  the case of the 6d conformal
% 4-derivative vector gauge  theory.  
 The  computation of $C_T$  for  2-derivative non gauge invariant  conformal  vector theory 
  in  generic $d$   (reducing to Maxwell theory for $d=4$) will  be discussed in Appendix \ref{app:erd}.
  
  \subsection{$\del^2$ gauge  vector in $d=4$}
  It  useful to start with recalling the computation of free  energy  of the Maxwell theory on 
  $S^1_q \times \H^3$. The  closely related case  of $S^1_q \times S^3$   background was discussed, e.g., in 
 section 2.2 of \cite{Beccaria:2014jxa}.  Starting  with $I= - {1\ov 4} 
 \int d^4x \sqrt g\, F^{\m\n}F_{\m\n}$  where $F_{\m\n}=\del_\m V_\n - \del_\n V_\m$   and  fixing the 
 $V_0=0$ gauge one  ends up with the  free energy expressed  in terms  of the operator defined on transverse 3-vector   (cf. \rf{4.2}) 
 %expressed in terms of the  determinant of the 
 %operator 
 \be 
\la{4.2x}
F_{q} = \tfrac{1}{2}\,\log {\det}\big(- \del_0^2  + {\bm \Delta_1}\big) \ , \qquad \qquad 
 {\bm \Delta_1} = - \DD^{2}-2  \ , 
\ee
 where $ {\bm \Delta_1} $  is the $d=4, \, s=1$ case of the operator in \rf{x2}
 with the  eigenvalue $\l$.  The corresponding  spectral density is the $s=1$   case of 
  \rf{5.9}, i.e. in $d=4$ and $d=6$  it reads
\be 
d\mu_{1, 3} = \tfrac{1+\lambda}{2\,\pi^{2}\,\sqrt\lambda}\,d\lambda, \ \qquad \ \ \ \qquad 
d\mu_{1, 5} = \tfrac{\sqrt\lambda\,(4+\lambda)}{6\,\pi^{3}}\,d\lambda  \ . \la{959}\ee
As a result,  the $\H^3$ part of heat kernel in \rf{x4}  is (cf. \rf{4.17}) 
\be \la{100}
\OK_{\H^3} = \int^\infty_0 d\lambda \tfrac{1+\lambda}{2\,\pi^{2}\,\sqrt\lambda}\,  e^{-t\l} 
= \tfrac{ 2 (1 + 2 t) }{(4 \pi t)^{3/2}} \ , 
\ee
and thus  integrating over $t$, dropping quartic and quadratic divergences   and summing over $n$ 
as in \rf{3535}   we get 
\begin{align}  \la{101}
&\F_{q} = \tfrac{1+30\,q^{2}}{180\,q^{3}}, \qquad\qquad \qquad 
\SS_{q} = -\tfrac{(1+q)(1+31\,q^{2})}{180\,q^{3}},
\\  &\la{102}
{\aa} = -\tfrac{1}{4}\,S_{1} = \tfrac{4}{45}\ ,\qquad  \qquad C_{T,4} =160 \,\cc=  80\,S'_{1} = 16 \ . 
\end{align}
This  reproduces the correct value of $C_{T}$ or   $\cc$-coefficient for the  Maxwell field
but not the standard  value of the a-coefficient that should be 
$ \frac{31}{180} = \frac{4}{45}+{\frac{1}{12}}$.  As  mentioned in section 2, this
 matching need not be expected  to follow automatically   when free energy is computed on 
   $S^1_q \times \H^3$    but 
one can  formally enforce   the relation between the 
$\SS_1$ and the   Weyl   anomaly $\aa$-coefficient 
  by shifting $\F_q$  and thus $\SS_q$    by a constant  as in \rf{2244}:
  \be \la{103} 
   \F_q \to  \F_q - \tfrac{1}{3} =  \tfrac{1+30\,q^{2}- 60 q^3}{180\,q^{3}},      \ \ \ \ \qquad  \ \   \ \ \ \ \   \SS_q \to  \SS_q - \tfrac{1}{3} 
  =  -\tfrac{1+q +31\,q^{2}  + 91 q^3}{180\,q^{3}} \ . 
\ee

\subsection{$\del^4$ gauge  vector in $d=6$}

 Defined on a curved background, the 6d    conformal   vector  gauge  theory  has  the following  
 Weyl-invariant action    \cite{Beccaria:2015uta} 
\be
\la{5.1}
I =  \int d^{6}x\,\sqrt{g}\,\big[D_{\lambda}F\indices{^{\lambda\mu}}
\,D^{\nu}F_{\nu\mu}-
\big(R_{\mu\nu}-\tfrac{1}{5}R\,g_{\mu\nu}\big)\,F^{\mu\lambda}\,F\indices{^{\nu}_{\lambda}}
\big],
\ee
where $F_{\mu\nu} = \del_\m V_\n - \del_\n V_\m $.
To compute the  corresponding free energy on $\Ss^1_q \times \H^{d-1}$ 
it is convenient to choose  again the temporal gauge $V_{0}=0$. This 
leads to  the  ghost factor $(\det \partial_{0}^{2})^{1/2}$ in the partition function. 
Using \rf{4444} the Lagrangian
in  (\ref{5.1})  then becomes (here $i,j, ...=1,...,5$ are  indices of $\mathbb H^{5}$)
\be\la{5.2} 
\mathscr L = (\del_0^2  V_i  +   D^k F_{ki}  )^{2} + (\del_0 D^i V_i)^{2}  -4\,(\partial_{0} V_{i})^{2}.
\ee
Change of   the variables $V_i\to (V_i^\perp, \varphi)$, 
\be\la{522}
V_{i} = V_{i}^{\perp}+\del_{i}\,\varphi, \qquad \qquad D^{i}\,V_{i}^{\perp}=0,
\ee
introduces the Jacobian factor $(\det \bm{D}^{2})^{1/2}$\  ($\DD^2\equiv D^iD_i$) 
 in the  path integral, 
while the Lagrangian \rf{5.2}  takes the form % becomes (using  \rf{4444})
\ba
\la{5.4}
\mathscr L &= \big[(-\partial_{0}^{2}+\bm{\Delta}'_{1})\,V^\perp_i\big]^{2}
-4\,(\partial_{0}V^{\perp}_{i})^{2}
+\varphi\,\partial_{0}^{2}\,\bm{D}^{2}\,(-\partial_{0}^{2}-\bm{D}^{2}-4)\,\varphi,
\\
\la{5.5}
& \bm{\Delta}'_{1} V^{\perp}_{i} \equiv  (-\bm{D}^{2}\,g_{ij}+R_{ij})\,
V^{\perp\, j} = (-\bm{D}^{2}-4)\,V^\perp_i =  (\bm{\Delta}_{1} +1) V^\perp_i \ , 
\end{align}
where $\bm{\Delta}_{1}$ is the $d=6$   case of the operator   defined in \rf{x2}. 
%is the  standard  transverse vector Laplacian.
 Integrating over $\varphi$
we get a factor $\big[\det(\partial_{0}^{2}\bm{D}^{2})\big]^{-1/2}$ (which cancels against  the 
 the previously  mentioned  
ghost and Jacobian factors) as well as
 the contribution  of  the  conformal 6d scalar  $\big[ \det 
(-\partial^{2}_{0}-\bm{D}^{2}-4) \big]^{-1/2}$   (see   \rf{4.2}). 

The   remaining  4-derivative operator acting on $V^\perp_i$  in \rf{5.4}  factorizes  exactly as   in 
the 4-derivative scalar case \rf{4.21}  (with $\bm{\Delta}_{0} \to \bm{\Delta}_{1}=\bm{\Delta}'_{1}-1  $)
\ba \la{5.6}
\OO^{(4)}_1= \big(-\partial_{0}^{2}+ \bm{\Delta}'_{1}\big)^{2}+4\,\partial_{0}^{2}
&=  \big[  (i \del_0 +1)^2 +  \bm{\Delta}_{1}   %  -   \DD^2   -\tfrac{(d-2)^{2}}{4} 
 \big]\,
  \big[  ( i\del_0 - 1)^2   +   \bm{\Delta}_{1}  \big]\, \\
  \la{5656}
 & =  \big[  (i \del_0)^2 + (\sqrt{ \bm{\Delta}_{1}} + i)^2   %  -   \DD^2   -\tfrac{(d-2)^{2}}{4} 
 \big]\,
  \big[  ( i\del_0 )^2   +  (\sqrt{ \bm{\Delta}_{1}} - i)^2  \big]\ . \end{align}
As in \rf{4.21},  % the same factorization is  found  on $S^1 \times S^5$.\footnote{
the same factorization  as in \rf{5.6},\rf{5656}  is  found  if one considers the theory \rf{5.1} on 
  $S^{1}\times S^{5}$.\foot{
  Due  to  the change of the sign of the curvature of the spatial part, 
  here $  \bm{\Delta}'_{1} = -\bm{D}^{2}+4$  that   has 
  discrete eigenvalues on the  sphere,  i.e. 
$\bm{\Delta}'_{1}  \to m^{2}+6m+8$ with integer $m\ge 0$.
The $4 \del_0^2$  term in   (\ref{5.6}) here  has  flippped  sign (as it came from the curvature term in \rf{5.1}) and   thus  we  find that on  $S^{1}\times S^{5}$
\begin{align}
\notag\te 
\big(-\partial_{0}^{2}+ \bm{\Delta}'_{1}\big)^{2}-4\partial_{0}^{2}  \ \to \ 
%\big(\frac{n^{2}}{q^{2}}+m^{2}+4m+4\big)\,\big(\frac{n^{2}}{q^{2}}
%+m^{2}+8m+16\big)\\
  \big[\big(\frac{n}{q}+i\big)^{2}+(m+3)^{2}\big]\,
\big[\big(\frac{n}{q}-i\big)^{2}+(m+3)^{2}\big]
=   \big[\frac{n^2}{q^2}+(m+2)^{2}\big]\,
\big[\frac{n^2}{q^2}+(m+4)^{2}\big]  \ .\notag
\end{align}
This  leads to  the  thermal free energy    corresponding to  the spectrum of dimensions 
$w_{m}= m+2, \ m+4$    
expected from the   operator   counting 
 on $\mathbb R\times S^{5}$ (as explicitly discussed 
in \cite{Beccaria:2014jxa}  in  the 4d  case). 
}
 Using that $\bm{\Delta}_{1}$ in \rf{x2}  has the eigenvalue $\l$, 
  the polynomial $P_\H$ in \rf{x3} is  again given  by \rf{417}. 
The  only   difference  compared  to  the 4-derivative 6d scalar case is that now 
the spectral measure is given  by the $s=1$ case of \rf{5.9}, i.e. by $d\mu_{1,5}$ in \rf{959}. 

As a result, the free energy  $\F_q$  and thus  the \renyi   entropy and $C_T$ 
for the 4-derivative  vector theory \rf{5.1} is given by the sum of the  contribution of the transverse 
spatial vector (with the  kinetic operator 
$\OO^{(4)}_1$)   and of the   standard  6d conformal scalar, i.e.
\be
\la{5.8}
\F_{q}(V^{(4)}) = 
\F_q(\bm{V}^{(4)}_\perp)+\F_q(\varphi) \ . 
\ee
 The   scalar   contribution   was already  given  in \rf{4217}--\rf{399}. 
 The total  $\H^5$  heat  kernel   factor in the resulting free   energy is then 
 (cf. \rf{y11}) 
 \be
\la{y112}
\OK_{\H^5}(t) = \int_{0}^{\infty}d\lambda\,\tfrac{\sqrt\lambda(4+\lambda)}{6\,\pi^{3}}
\,e^{-t\,\lambda} \big[ 1  +   2 \, e^{t}   \cos( 2 t\sql) \big] 
=  \tfrac{ 9 - 10 t - 32 t^2}{\,(4\,\pi\,t)^{5/2}}\ ,\ee
and thus   finally 
\begin{align} 
&\F_{q} = \tfrac{-6+35\,q^{2}+1680\,q^{4}}{10080\,q^{5}}, \qquad\qquad
\SS_{q} = -\tfrac{(1+q)(-6+29\,q^{2}+1709\,q^{4})}{10080\,q^{5}},\la{y22}\\
&
{\aa} = -\tfrac{1}{96}\,\SS_{1} = \tfrac{433}{24\times 7!}\ , \qquad \qquad 
C_{T,6} =3024 \cc_3= -504\,\SS'_{1} = -90\ . \la{y222}
\end{align}
The value  of $C_{T,6} $  is the same \rf{115} as quoted in  the Introduction, 
  found earlier  by  other methods 
in  \cite{Osborn:2015rna,Giombi:2016fct}.
To  also  reproduce  the correct value $\aa= {275\ov 8\times 7!}$   \ci{Beccaria:2015uta}
of the a-anomaly for the 4-derivative 6d vector field  one needs,  as in the $d=4$ vector case  \rf{103}, 
   to shift 
$\F_q$ and thus $\SS_q$    by  the   constant  term  \ $-{14\ov 45}$.

%%%%%%%%%%%%%%%%%%%%%%%%%%%%%%%%%%%%%
\section{Fermionic fields}

Finally,  let us  discuss  the fermionic fields. 
We shall first review  the computation of  free energy and $C_T$ for the standard   Dirac fermion 
and then  consider the  conformal 3-derivative fermion  which is part of the 6d superconformal vector multiplet \rf{1.10}.

\subsection{$\slashed{\partial}$ fermion}

The curved space  Weyl-invariant action for  a  standard  massless fermion 
\be
I =i \int d^{d}x\,\sqrt{g}\,\overline\psi\,\slashed{D}\,\psi,
\ee
leads to the following  formal expression for its  
  free energy on ${S^{1}_q\times \mathbb H^{d-1}}$
 in terms of the  eigenvalues of the squared operator  $(i\slashed{D})^{2}= -\del_0^2 + (i\slashed{\bm{D}})^{2}  $ 
\cite{Camporesi:1992tm,Camporesi:1995fb}
\be
\la{63}
\mc F_{q} =-\text{tr}\,\log(i\,\slashed{D})=
 -\ha n_{f}\,{\VV_{\mathbb H^{d-1}}}
 \sum_{n\in\mathbb Z+\frac{1}{2}}
\int_{0}^{\infty}d\mu_{\frac{1}{2}, d-1}(\lambda)\,\log\big(\frac{n^{2}}{q^{2}}+\lambda\big) \ .
\ee
Here $n_{f}$  is the complex 
dimension of the spinor space  (e.g., $n_f=2$ for a  Weyl fermion in $d=4$ or 
MW fermion in $d=6$)
and the sum over  half-integer $n$  corresponds to the antiperiodic boundary conditions on 
the "thermal" circle  $S^{1}$.
 $\l$ is the  eigenvalue of the operator 
$(i\slashed{\bm{D}})^{2} = - \bm{D}^2 + {1\ov 4} R $  equal to 
$ - \bm{D}^2 - {1\ov 4} (d-1)(d-2) $ on   $\H^{d-1}$ (cf. \rf{x2}).
 The corresponding  spin 1/2  Plancherel measure for even $d$ is \cite{Camporesi:1992tm,Camporesi:1995fb} (cf. \rf{5.9})
\begin{align}\la{6.3}
&d\mu_{\frac{1}{2}, d-1} = 
\frac{1}{2^{d-2}\,\pi^{d-1\ov 2}\,\Gamma(\frac{d-1}{2})}\left|\frac{\Gamma(i\,\sql+\frac{d-1}{2})}{\Gamma(i\,\sql
+\frac{1}{2})}\right|^{2}\,d\sql \ , \\
&\la{633}
d\mu_{\frac{1}{2}, 3} = \tfrac{1+4\,\lambda}{16\,\pi^{2}\,\sqrt\lambda}\,d\lambda, 
\qquad\qquad 
d\mu_{\frac{1}{2}, 5} = 
\tfrac{(1+4\,\lambda)(9+4\,\lambda)}{384\,\pi^{3}\,\sqrt\lambda}\,d\lambda, \quad ... \ . 
\end{align}
Using the proper time representation for the  log in \rf{63} 
we get as in \rf{x4},\rf{x7} 
\begin{align} 
\la{x4f}
&\qquad \qquad \F_{q} =  %\sum_{k=1}^p \F_q^{(k)} \ , \ \ \ \ \qquad \quad  \F_q^{(k)} =
 -  \ha n_f \, {\VV_{\mathbb H^{d-1}}} \, 
  \int ^\infty_0 {dt \ov t}  \, K^f_{S^1}(t) \, \overline  K_{\H^{d-1}}(t) 
%\int^\infty_0 d\mu(\lambda)\,    e^{- t  (\sql + i  \a_{k})^{2} }
  \ ,\\
& \la{x5f}   K^f_{S^1}(t) = \tfrac{2\,\pi\,q}{(4\,\pi\,t)^{1/2}}\sum_{n\in\mathbb Z}  (-1)^n  \, e^{ 
-\frac{\pi^{2}\,n^{2}\,q^{2}}{t}}   \ , \qquad \qquad 
\overline  K_{\H^{d-1}}(t) = \int^\infty_0  d\mu_{\frac{1}{2}, d-1}(\lambda)\,   e^{- t \, \l } \ . 
\end{align}
The power UV divergences  in the  proper time integral 
should  be again subtracted by omitting the  $n=0$ term in the sum. 
%over $n$. 
Explicitly, one finds in $d=4$  
\ba\la{7667}
&\OK_{\H^3}(t) = \int_{0}^{\infty}d\lambda\,\tfrac{1 + 4 \l  }{4\,\pi^{2} \sqrt\lambda}
\,e^{-t\,\l}\, 
=  \tfrac{2+\,t}{2(4\,\pi\,t)^{3/2}}\ , \\
\la{5757}
&
\mc F_{q} = \tfrac{7+30\,q^{2}}{2880\,q^{3}}\,n_{f}\ ,\qquad\qquad
\SS_{q} = -\tfrac{(1+q)(7+37\,q^{2})}{2880\,q^{3}}\,n_{f}
\ , \\
& {\rm a} = -\tfrac{1}{4}\,S_{1} = \tfrac{11}{1440}\,n_{f}\ , \qquad \qquad 
C_{T,4} =160 \, \cc= 80\,\SS'_{1} = 
2\,n_{f}\ , \la{577}
\end{align}
and in $d=6$
\ba
\la{5767}
&\OK_{\H^5}(t) =  \int_{0}^{\infty}d\lambda\,\tfrac{\sqrt\lambda(1+4\lambda)(9+4\l)}{384\,\pi^{3}\sql }
\,e^{-t\,\lambda}
=  \tfrac{ 2-{ 10\ov 3}  t + {3\ov 2} t^2 }{2(4\,\pi\,t)^{5/2}}\ , \\
\la{7575}
&\mc F_{q} = -\tfrac{31+245\,q^{2}+945\,q^{4}}{483840\,q^{5}}\,n_{f}\ ,\qquad\qquad
\SS_{q} =\tfrac{(1+q)(31+276\,q^{2}+1221\,q^{4})}{483840\,q^{5}}\,n_{f} \ ,
\\
&\la{751}
{\rm a} = -\tfrac{1}{96}\,S_{1} = -\tfrac{191}{576\times 7!}\,n_{f}\ , \qquad \qquad C_{T,6} = 
3024\,  \cc_3=
-504\,S'_{1} = 
3\,n_{f}\  . 
\end{align}
Eqs. \rf{577},\rf{751} give  the correct known  values of the  $\aa$ and $\cc$  Weyl anomaly 
coefficients in $d=4$ \ci{Duff:1977ay}  and in $d=6$  \ci{Bastianelli:2000hi} and the 
values of $C_{T}$  also 
agree   with the general  expression  for $C_{T,d}(\psi) $ given in  (\ref{1.16}).
The expressions  for the \renyi entropy agree with \ci{Klebanov:2011uf,Nian:2015xky}.

%%%%%%%%%%%%%%%%%%%
\subsection{Conformal $\slashed{\partial}^{3}$ fermion}

The Weyl-invariant  operator for a  3-derivative 
 fermion was first   found    in the context  of  extended 
 conformal supergravity  \cite{Bergshoeff:1980is}  
 in  $d=4$ \ci{Fradkin:1981jc}  (for Majorana  fermions) 
  \be
\la{6.8}
I = i \int d^{4}x\,\sqrt{g}\,\overline\psi\, \big[   \slashed{D}^{3}+\big(R^{\mu\nu}-\tfrac{1}{6}R g^{\m\n}\big)\,\gamma_{\mu}\,D_{\nu}   \big]\,\psi \ . 
\ee
In $d=6$ the  analogous   3-derivative  operator was recently  found in  \cite{Butter:2017jqu}\foot{We thank  D. Butter for pointing this out to us and a clarifying discussion.} 
%The six dimensional case has been discussed recently in \cite{Butter:2017jqu} and takes a similar 
%form to (\ref{6.8}), but with different coefficients and an extra term
\be
\la{6.9}
%d=6:\qquad \qquad
 \OO^{(3)} = \slashed{D}^{3}+\tfrac{1}{2}\big(R^{\mu\nu}-\tfrac{1}{10}Rg^{\m\n} \big)\,\gamma_{\mu}\,D_{\nu}+\tfrac{1}{10}\,\gamma^{\mu} D_{\mu}R\,.
\ee
%The extra term is zero in 4d because for Majorana spinors $\overline\psi\gamma_{\mu}\psi=0$. 
The generalization of \rf{6.8},\rf{6.9}  to any $d$   reads %($P_{\m\n} $ is the Schouten tensor   as in \rf{999})
\ba\la{699}
\OO^{(3)} = \slashed{D}^{3}+  2 P^{\m\n} \, \gamma_{\mu}\,D_{\nu}   +    \,\gamma^{\mu} D_{\mu} P \  ,  %, \\
%P_{\m\n} &= \tfrac{1}{d-2}\,\big(R_{\mu\nu}-\tfrac{1}{2(d-1)} R\big) \ , \qquad \qquad 
%P= P^\m_\m = \tfrac{1}{2(d-1)} R \ . \la{9999}
\end{align}
where $P_{\m\n} $ is the Schouten tensor   as in \rf{999}. 
On $S^{1}\times \mathbb{H}^{d-1}$ we   can use \rf{4444}  to get the following explicit form of \rf{699} 
\be\la{1000}
\OO^{(3)} = \slashed{D}^{3}+\slashed{D}-2\,\slashed{\bm{D}} = (\gamma^{0}\partial_{0}+
\slashed{\bm{D}})^{3}+\gamma^{0}\partial_{0}-\slashed{\bm{D}}.
\ee
%%%%%%%%%%%%%%%%%%%
%%%%%%%%%%%%%%%%%%%%%%
As a result,  its square factorizes   in a $d$-independent  manner  just   like in the 4-derivative scalar  
case in  \rf{4.21} (cf. also \rf{5.6})\footnote{The factorization of the operator  (\ref{6.8}) on $S^{1}\times S^{3}$  was  observed   in \cite{Beccaria:2014xda}.}
\ba
(i \OO^{(3)})^{2}  &= - \del_0^2\,   (\partial_{0}^{2}+\slashed{\bm{D}}^{2} + 1 )^{2}-
\slashed{\bm{D}}^{2}\,(\partial_{0}^{2}  + \slashed{\bm{D}}^{2}-1)^{2}
\no \\
&\la{6.12}
=(-\del_0^2 - \slashed{\bm{D}}^{2})\big[  (i\del_0 +1)^2 - \slashed{\bm{D}}^{2}\big]
\big[ (i\del_0-1)^2 - \slashed{\bm{D}}^{2} \big] \\
&\la{6123}
=(-\del_0^2 - \slashed{\bm{D}}^{2})\big[  (i\del_0)^2  + (i\slashed{\bm{D}}  + i)^{2}\big]
\big[ (i\del_0)^2 + ( i  \slashed{\bm{D}} - i)^{2} \big]
\ . \end{align}
The corresponding eigenvalue polynomial  \rf{x3} is then
 the product of the standard  fermion part  and the  same  factor as in the $\del^4$ scalar  case  in \rf{417} (and also has a similar structure as the result in the 4-derivative vector case in \rf{5656}). 

Using the expression for the spin  1/2  spectral measure in \rf{633}   and starting with \rf{x4f} 
 we then find in $d=4$  (cf. \rf{17}--\rf{711}) 
\begin{align}
\la{18}
&\OK_{\H^3}(t) = \int_{0}^{\infty}d\lambda\,\tfrac{1 + 4 \l  }{4\,\pi^{2} \sqrt\lambda}
\,e^{-t\,\lambda}\, \big[1 + 2\, e^t   \cos( 2 t\sql) \big] 
=  \tfrac{6- 5\,t}{2(4\,\pi\,t)^{3/2}}\ , \\
&\mc F_{q} =  \tfrac{7-50\,q^{2}}{960\,q^{3}}\, n_f \ , \qquad\qquad 
\SS_{q} =  \tfrac{(1+q)(-7+43\,q^{2})}{960\,q^{3}}\, n_f\ ,\la{81}\\
&{\rm a} = -\tfrac{1}{4}\,\SS_{1} = -\tfrac{3}{160}n_f \ , \qquad 
\qquad C_{T,4} = 160 \, \cc= 80\,\SS'_{1} = -\tfrac{2}{3}n_f \ .\la{811}
\end{align}
The  values of $\aa=- {3\ov 80}$ and $\cc= -{1\ov 120}$  for a 
Majorana fermion ($n_f=2$)   agree with the ones found 
by direct computation in \ci{Fradkin:1981jc,Fradkin:1985am}.\foot{In  the notation of Table 6.1  in  
\ci{Fradkin:1985am}  one  has for the 3-derivative  $\Lambda$-spinor: 
 $\b_1= {7\ov 240}, \  \b_2 =  -{1\ov 60} $    with    
 $\aa= -\b_1 + \ha \beta_2, \ \cc= \ha \beta_2$.
 To compare, for a real 4-derivative scalar
  $\b_1= {1\ov 90} \ , \ \b_2=- { 2 \ov 15} $,  giving $\aa= -{7\ov 90}  , \  \cc= -{1\ov 15}$ 
  in agreement with values given  earlier  in \rf{711} (note that the $\del^4$  scalar $\vp$ in Table 6.1 is complex).  }
  
In 6  dimensions  we get   (cf. \rf{y1}--\rf{y3})  and \rf{5767}--\rf{751}) 
\begin{align}
\la{y19}
&\OK_{\H^5}(t) =  \int_{0}^{\infty}d\lambda\,\tfrac{\sqrt\lambda(1+4\lambda)(9+4\l)}{384\,\pi^{3}\sql }
\,e^{-t\,\lambda} \big[1 + 2\, e^t  \cos( 2 t\sql) \big]
=  \tfrac{ 3-3t -t^2 }{(4\,\pi\,t)^{5/2}}\ , \\
&
\F_{q} = \tfrac{-31+147\,q^{2}+735\,q^{4}}{161280\,q^{5}}\,n_{f}\ ,\qquad\qquad 
\SS_{q} = -\tfrac{(1+q)(-31+116\,q^{2}+851\,q^{4})}{161280\,q^{5}}\,n_{f}, \la{y20}\\
&
{\aa} = -\tfrac{1}{96}\,S_{1} = \tfrac{39}{64\times 7!}\,n_{f} \ ,\qquad  \qquad C_{T,6} 
= 3024\, \cc_3=-504\,S'_{1} = 
-\tfrac{18}{5}\,n_{f}
\ .\la{y39}
\end{align} 
Thus  for  a 6d MW   fermion with $n_f=2$  we get $\aa= \tfrac{39}{32\times 7!}$ in agreement 
with the value found in  \ci{Beccaria:2015uta}
while 
\be  C_{T,6}(\psi^{(3)}) = -\tfrac{36}{5}  \ . \la{y21}\ee
This  confirms  the value corresponding to $\xiyz$ in \rf{118}.

To emphasize that \rf{y21} is a result of a rather  non-trivial  computation,  
in Appendix \ref{app1}  we shall present an alternative  way of arriving at \rf{y21} 
based on the approach that does not use  the proper time representation 
and utilizes the first way \rf{6.12} of  factorizing the square of the 3-derivative spinor 
operator \rf{1000}. Surprisingly, a  naive application of this alternative approach   
leads precisely to the  value of $C_{T,6}$ in \rf{1180} corresponding to $\xibt$  in \rf{1.8}.

It is possible to generalize the $d=4$  \rf{811}  and $d=6$  \rf{y39}
 expressions for  $C_T$ of  the 3-derivative    conformal fermion    to any  dimension 
$d$ obtaining the following counterpart  of the general $d$ expressions 
for $C_T$ of  the standard scalar and spinor  \rf{1.16},   4-derivative scalar \rf{444}  and 6-derivative scalar \rf{a1}
\be\la{6156}
\te C_{T,d}(\psi^{(3)}) =    - n_{f} \, \frac{d (d^2 + d -18) }{2(d-1)(d-2)}
 = -   \frac{d^2 + d -18 }{(d-1)(d-2)}  \ C_{T,d}(\psi)   \ .
\ee
It would   be interesting to reproduce  \rf{6156}  in   alternative  flat-space approaches like  the ones used in  the higher-derivative scalar cases  in   \cite{Guerrieri:2016whh}  and \ci{Osborn:2016bev}.

%%%%%%%%%%%%%%%%%%%%%%%%%%%%%%
\section{Conformal  anomaly  of   6d
  higher derivative vector  multiplet   from\\ Seeley - DeWitt  coefficient} %4-derivative  conformal  6d  gauge vector }
%%%%%%%%%%%%%%%%%%%%%%%%%%%%%%%%%%%%
\la{sec:seeley}

Let us now rederive   the above results  for the  $\cc_3$  coefficient in \rf{1.1}   for the fields  of the vector multiplet    \rf{1.10} 
%that agree with \rf{188} 
by  the same direct method as used in \ci{Bastianelli:2000hi}   to  compute  the conformal   anomalies of the 
standard  fields in the  (2,0)  tensor multiplet -- using the  general expression \ci{Gilkey:1975iq} 
for the $b_6= \langle T^\m_\m\rangle$   Seeley-DeWitt coefficient of the 2nd order Laplace-type operator $\Delta= - D^2  + X $.
%\foot{Below 
%we shall follow the notation   in  \ci{Bastianelli:2000hi}    with $  b_6  (4\pi)^{3} \langle T^\m_\m\rangle

 The two key observations that allow  one to do this are:\foot{We are  grateful to D. Diaz   for    suggesting this approach to us.}
   (i) like the  higher derivative  conformal scalar operators \ci{Gover:2005mn},  
    the 4-derivative vector  operator  in  \rf{5.1} 
 and  the square of the 3-derivative spinor operator in \rf{6.9}  factorize  into  a product of  2nd order Laplacians 
 on an Einstein space  $R_{\m\n} = {1\ov 6} R g_{\m\n}$   and thus
  their anomalies can be readily computed;   (ii) considering a   general Einstein 
  background  is sufficient  to determine  all the 4 anomaly coefficients $\aa,\ \cc_i$ in \rf{1.1}.
  The special cases       were   already    considered  before -- 
   the 6-sphere (allowing to find   the $\aa$-coefficient \ci{Beccaria:2015uta})  and the Ricci-flat space (allowing  to fix the  $\cc_i$  up to one free parameter \ci{Beccaria:2015ypa}).  On a general Einstein background   one may have both the  scalar curvature  and  the Weyl 
   tensor non-zero  so that one may  capture the    $ R C^{\m\n\l\r} C_{\m\n\l\r}$  terms  in 
   the  expression for  $b_6$  and thus  determine 
     one more 
   combination of the Weyl anomaly coefficients.  
   
   As a result,  in addition to the  anomaly coefficients   for the conformally coupled 2-derivative  6d scalar 
   ($\Delta= - D^2  + {1 \ov 5} R$)   
   obtained  in \ci{Bastianelli:2000hi} 
   \be\la{7.1}
  % \del^2 - {\rm scalar}: \ \ \ \ \ 
 \te   \aa= - { 5 \ov 72\times 7! } \ , \ \quad  \ \    \cc_1=- { 28 \ov  3\times  7!}  \ , \quad \ \   \cc_2= { 5 \ov 3 \times 7! } \ , \ \quad  \   \cc_3= { 2 \ov 7! }\ , 
   \ee
   we find  for the 4-derivative conformal vector  \rf{5.1}
       \be\la{7.2}
  % \del^4 - {\rm  vector}: \ \ \ \ \ 
 \te   \aa=  { 275 \ov 8\times 7!} \ , \ \quad \ \ \ \   \cc_1= { 2716 \ov  7! }  \ , \ \quad \   \cc_2= { 911 \ov  7! } \ , \ \ \quad    \cc_3= - { 150 \ov 7! }\ , 
   \ee
   and 3-derivative conformal MW spinor 
  \be\la{7.3}
  % \del^4 - {\rm  vector}: \ \ \ \ \ 
 \te   \aa=  { 39 \ov  32\times 7!} \ , \quad \ \ \ \ \   \cc_1={ 448 \ov 3 \times 7! }  \ , \quad \ \   \cc_2= { 110 \ov 3\times 7! } \ , \quad \ \   \cc_3= -{ 12 \ov 7! }\ . 
   \ee
   As a result, the anomaly   coefficients  for  
    the  higher derivative vector  multiplet   \rf{1.10}   are found to be\foot{This   
    corrects the expressions for $\cc_i$  in  eq. (2.3) in  \ci{Beccaria:2015ypa}   that assumed the 
   wrong value of $\xi$ in \rf{1.8}.} 
    \be\la{7.4}
  % \del^4 - {\rm  vector}: \ \ \ \ \ 
 \te   \aa=  { 1757 \ov  48\times 7!} \ , \quad \ \ \ \ \   \cc_1= {8960 \ov 3 \times 7! }  \ , \quad \ \   \cc_2= { 2968 \ov 3\times 7! } \ , \ \ \quad  \cc_3= -{ 168 \ov 7! }\ . 
   \ee
   The values of $\aa$-coefficient   were found   already in the special case of  $S^6$ background in  \ci{Beccaria:2015uta}.
   The  (1,0)   supersymmetry constraint  $\cc_1- 2 \cc_2 + 6 \cc_3=0$   in  \rf{1.2} 
   and   the relation $\cc_1 + 4 \cc_2 = {62\ov 45}$ were obtained  by considering  the Ricci-flat background in \ci{Beccaria:2015ypa}. 
   The coefficients $\cc_3$     in \rf{7.2}   and \rf{7.3}  (or $C_{T,6}= 3024 \cc_3$ in \rf{1.13})
   are the same as  the ones found above  in \rf{y222}  and  \rf{y39} from the  computation of free energy on $S^1\times \H^5$.
   The values   of $\cc_i$   in \rf{7.4}  thus   agree 
   %are consistent with the (1,0)   supersymmetry constraint  $\cc_1- 2 \cc_2 + 6 \cc_3=0$   in  \rf{1.2} 
    with \rf{1.11}  for the   value of $\xi=-{8\ov 9}$   found in \ci{Yankielowicz:2017xkf}
     providing another independent confirmation  of  \rf{188}.
     
    %v3-at
    In Appendix \ref{app:Vp}  we shall present the extension of the computation presented in this section to 
more general $(1,0)$ superconformal multiplets with maximal spin 1  with the results that are again in agreement with \rf{1.7} with \rf{188}.

    \def\bb {\bar b}\def \D {\Delta} \def \d {\delta} 
    
    Below we shall  follow  the notation in  \ci{Bastianelli:2000hi}    and use    that for an Einstein   background one has 
     $D_\m R=0$  and $D^\m C_{\m\n\l\r}=0$  so that many   terms in  the general expression in $b_6$  simplify. 
     The $E_{6}$ and $I_{1,2,3}$ invariants  in \rf{1.1} defined in   \ci{Bastianelli:2000hi}   take the form\foot{
     The only non zero total derivative terms  among  $C_{1,\dots, 7}$  in   \ci{Bastianelli:2000hi}  here are
\begin{align}\no 
C_{5} &= C^{\alpha\beta\gamma\delta} D^{2}C_{\alpha\beta\gamma\delta} +
(D_{\mu}C_{\alpha\beta\gamma\delta})^{2} = 
D_{\mu}(C_{\alpha\beta\gamma\delta} D^{\mu}C^{\alpha\beta\gamma\delta}), \notag \\
C_{7}&=\tfrac{1}{12} R C_{\alpha\beta\gamma\delta} C^{\alpha\beta\gamma\delta} -  \
C_{\alpha}{}^{\mu}{}_{\gamma}{}^{\nu} C^{\alpha\beta\gamma\delta} \
C_{\beta\mu\delta\nu} -  \tfrac{1}{4} C_{\alpha\beta}{}^{\mu\nu} \
C^{\alpha\beta\gamma\delta} C_{\gamma\delta\mu\nu} 
 + \tfrac{1}{4} (D_{\mu}C_{\alpha\beta\gamma\delta})^{2}.\no 
\end{align}
  } 
\begin{align}
E_{6} &= - \tfrac{16}{75} R^3 -  \tfrac{8}{5} R C_{\alpha\beta\gamma\delta} \
C^{\alpha\beta\gamma\delta} + 64 C_{\alpha}{}^{\mu}{}_{\gamma}{}^{\nu} \
C^{\alpha\beta\gamma\delta} C_{\beta \mu \delta \nu} - 32 C_{\alpha\beta}{}^{\mu\nu}
 C^{\alpha\beta\gamma\delta} C_{\gamma\delta \mu\nu}, \notag \\
I_{1} &= - C_{\alpha}{}^{\mu}{}_{\gamma}{}^{\nu} C^{\alpha\beta\gamma\delta} \
C_{\beta \mu \delta \nu}, \qquad \qquad 
I_{2} = C_{\alpha\beta}{}^{\mu\nu} C^{\alpha\beta\gamma\delta} C_{\gamma\delta
\mu\nu}, \la{PPPP} \\
I_{3} &= - \tfrac{6}{5} R C_{\alpha\beta\gamma\delta} C^{\alpha\beta\gamma\delta} + 8 \
C_{\alpha}{}^{\mu}{}_{\gamma}{}^{\nu} C^{\alpha\beta\gamma\delta} \
C_{\beta\mu\delta\nu} + 2 C_{\alpha\beta}{}^{\mu\nu} C^{\alpha\beta\gamma\delta} \
C_{\gamma\delta\mu\nu} \notag \\ 
&\qquad  + 6 C^{\alpha\beta\gamma\delta} \,D^{2}\,C_{\alpha\beta\gamma\delta} + 3 \
(D_{\mu}\,C_{\alpha\beta\gamma\delta})^{2}.\no
\end{align}
Given a  general scalar  Laplacian 
    \be \la{PP} \te 
    \Delta_0(\k) \equiv -D^2 + \k \bar R, \ \ \qquad \qquad  \bar R = {1\ov d (d-1)} R = {1\ov 30} R \ , \ee
    the  corresponding $b_6$   coefficient computed as in  \ci{Bastianelli:2000hi}  is found to be 
    ($\bb_6 \equiv (4\pi)^{3}b_{6}$)
    \begin{align}
\la{P.1}
  7!\, \bb_{6} \big[ \D_0(\k)\big]    = &(\tfrac{2278}{675} + \tfrac{56}{25} \kappa + \tfrac{7}{15}  \kappa^2 + \tfrac{7}{225}  \kappa^3) \, R^3
\notag \\ 
& + (\tfrac{194}{45} + \tfrac{14}{ 15} \kappa) R\, C_{\a\b\g\d} C^{\a\b\g\d} \
+ \tfrac{80}{9} C_{\a}{}^{\m}{}_{\g}{}^{\n} C^{\a\b\g\d} \, C_{\b\m\d\n} \notag \\ 
& + \tfrac{44}{9} C_{\a\b}{}^{\m\n} C^{\a\b\g\d} \, C_{\g\d\m\n} 
+ 12 C^{\a\b\g\d} D^{2}C_{\a\b\g\d} 
 + 9 (D_{\m}C_{\a\b\g\d})^{2}.
\end{align}
Expressing this in terms of the invariants in \rf{1.1} using \rf{PPPP}  and ignoring the total derivative terms   we find that 
in the  special case  of the conformally  coupled  scalar  when $\k= {1 \ov 4}  d(d-2)=6$  we reproduce the  coefficients in \rf{7.1}. 

\def \F  {{\cal F}}

The 4-derivative vector  operator  in \rf{5.1} restricted to   an  Einstein background  factorizes  in the same  way as  in the sphere 
case discussed in \ci{Beccaria:2015uta}:   the action depends  only on
the  transverse part $V^\perp_\m$ of the vector 
 and   reduces to the integral of  $V^\perp  \Delta_{1\perp}(7)\,\Delta_{1\perp}(5)  V^\perp$. The
  resulting  partition function is then given by 
\begin{align}
%Z(V^{(2)}) &= \bigg[\frac{1}{\det\Delta_{1\perp}(7)\,\det\Delta_{0}(6)}\bigg]^{1/2} = 
%\bigg[\frac{\det\Delta_{0}(2)}{\det\Delta_{1}(7)\,\det\Delta_{0}(6)}\bigg]^{1/2}, \notag \\
Z(V^{(4)}) = \bigg[\frac{\det\Delta_{0}(0)}{\det\Delta_{1\perp}(7)\,\det\Delta_{1\perp}(5)}
\bigg]^{1/2} = 
\bigg[\frac{\det\Delta_{0}(2)\,\big[\det\Delta_{0}(0)\big]^{2}}
{\det\Delta_{1}(7)\,\det\Delta_{1}(5)}\bigg]^{1/2} \ , \la{P2} 
\end{align}
where like in \rf{PP}   we defined 
 $\Delta_1(\k)  V_\m  \equiv  (- D^2 + \k \bar R) V_\m$ and $ \Delta_{1\perp}$ is $\Delta_1$  restricted to 
$V^\perp_\m$.  The standard vector Laplacian on an Einstein background 
  is $(-D^2 g_{\m\n} + R_{\m\n}) V^\n= \Delta_1(5) V_\m$. 
In \rf{P2}  we used that 
$\det  \Delta_{1}(\k) = \det  \Delta_{1\perp }(\k)\, \det \Delta_0(\k-5)$.\foot{If $V_\m = V^\perp_\m + \del_\m \vp$, then 
on an Einstein background one has  (dropping total derivatives) 
$V^\m  \Delta_{1}(\k) V_\m = V^{\m\perp}  \Delta_{1\perp}(\k) V^\perp_\m + \vp \Delta_0(\k-5) \vp$.}  
To find the  $b_6$  coefficient   for the vector Laplacian  $\Delta_1(\k)$  from the general  expressions in 
  \ci{Gilkey:1975iq,Bastianelli:2000hi}
 one is to note that here  
the  covariant   derivative contains the extra vector connection part with the  "internal"
 curvature $(\F_{ij})_{\a}^{\ \ \b}=R\indices{_{ij\a}^{\b}}$. The analog of \rf{P.1} then reads
\begin{align}
\la{P.2}
 7!\,\bb_{6} \big[ \D_1(\k)\big] 
 = &(\tfrac{3394}{225} + \tfrac{938}{75} \kappa + \tfrac{14}{5} \kappa^2 + \tfrac{14}{75} \kappa^3) \,
R^3 
\notag \\ 
&     -  \tfrac{36}{5} (6 + \tfrac{7}{6} \kappa)    R\, C_{\a\b\g\d} C^{\a\b\g\d} \
+ \tfrac{80}{9} C_{\a}{}^{\m}{}_{\g}{}^{\n} C^{\a\b\g\d} \, C_{\b\m\d\n} \notag \\ 
& -  \tfrac{164}{3} C_{\a\b}{}^{\m\n} C^{\a\b\g\d} \, C_{\g\d\m\n} 
-96 C^{\a\b\g\d} D^{2}C_{\a\b\g\d} 
 -58 (D_{\m}C_{\a\b\g\d})^{2}.
\end{align}
This  generalizes  the expression  found in \ci{Bastianelli:2000hi}
in the special  case of the  standard vector Laplacian (corresponding to $\k=5$).

Eqs.  \rf{P.1}  and \rf{P.2}  are  all we need  to compute the anomalies of the 4-derivative conformal vector 
 since according to \rf{P2} 
\be \la{P3}
b_{6}(V^{(4)}) = b_{6} \big[ \D_1(7)\big]   + b_{6} \big[ \D_1(5)\big]   
- b_{6} \big[ \D_0(2)\big]   - 2 b_{6} \big[ \D_0(0)\big]  \ . 
\ee
As a result,  one finds the coefficients  given in \rf{7.2}. 

The 3-derivative  conformal spinor operator \rf{6.9}  restricted to an  Einstein   background becomes 
$\mc O^{(3)} =  \slashed{D}^3 + \tfrac{1}{30} R  \slashed{D} $  so that its square factorizes as 
\be\la{P4}
\big(i \mc O^{(3)}\big)^2 =  (-D^{2}+\tfrac{1}{4}\,R)\,(-D^{2}+\tfrac{13}{60}\,R)^{2} = \D_{1\ov 2} ( \tfrac{15}{2}) \ \big[ \D_{1\ov 2} ( \tfrac{13}{2})\big]^2  \ , 
\ee
where $\D_{1\ov 2} ( \k) \equiv  - D^2  + \k \bar R$    acting on spinors 
 has  $D$    being spinor  covariant derivative  with  the corresponding "internal" curvature $\F_{ij} = \tfrac{1}{4}\,R_{ijab}\,\gamma^{ab}$. 
The counterpart  of \rf{P.1} and \rf{P.2} in the spinor  case is  then  found to be\foot{Here we included the -1  fermion sign   factor.  $n_f$ is the  complex  dimension of the  spinor space  as in \rf{1.16}  equal to 2  in the 6d MW spinor case.} 
\begin{align}
\la{P.3}
 7! ( n_f)^{-1} \bb_{6} \big[ \D_{1\ov 2} (\k)\big] 
 = &- ( \tfrac{7369}{2700} +  \tfrac{637}{300} \kappa +   \tfrac{7}{15}  \kappa^2 +  \tfrac{7}{225}   \kappa^3) \,
R^3 
\notag \\ 
&     + (\tfrac{389}{90} + \tfrac{49}{60} \kappa)    R\, C_{\a\b\g\d} C^{\a\b\g\d} \
+ \tfrac{109}{9} C_{\a}{}^{\m}{}_{\g}{}^{\n} C^{\a\b\g\d} \, C_{\b\m\d\n} \notag \\ 
& + \tfrac{101}{18}   C_{\a\b}{}^{\m\n} C^{\a\b\g\d} \, C_{\g\d\m\n} 
+9 C^{\a\b\g\d} D^{2}C_{\a\b\g\d} 
 +5 (D_{\m}C_{\a\b\g\d})^{2}.
\end{align}
For  the standard  spinor  field   with the squared operator  being $\D_{1\ov 2} ( \k)$  with $\k= {15\ov2}$ 
eq. \rf{P.3}   reproduces   the    coefficients in \rf{1.1}  found in \ci{Bastianelli:2000hi}, 
i.e. for MW  spinor  with $n_f=2$ we get 
$ \aa=  -{ 191 \ov  288\times 7!} \ , \ \ \  \cc_1=-{ 224\ov 3 \times 7! }   , \ \ \   \cc_2=- { 8\ov  7! }  , \ \ \   \cc_3= { 10\ov 7! }\
$.
Using  that  for  the 3-derivative  spinor  we have from \rf{P4} 
$  b_6 (\psi^{(3)}) = b_{6} \big[ \D_{1\ov 2} ( \tfrac{15}{2}  )\big]  +  b_{6} \big[ \D_{1\ov 2} ( \tfrac{13}{2}  )\big]
$, we  find that the corresponding  conformal anomaly coefficients are given  by \rf{7.3}.

%%%%%%%%%%%%%%%%%%%%%%%%%%%%%%%%%%%%%%%%%%%%%%%%%
\section*{Acknowledgments}
%%%%%%%%%%%%%%%%%%%%%%%%%%%%%%%%%%%%%%%%%%%%%%%%%%
We would like  to thank D. Butter,   C.-M. Chang,  D. Diaz,  S. Giombi, Y.-H. Lin,  H. Osborn,  
A. Petkou, S. Solodukhin,   S. Yankielowicz  and Y. Zhou    for useful  discussions and comments. 
The work of AAT  was   supported by the ERC Advanced grant no. 290456,
 the  STFC Consolidated grant ST/L00044X/1, by the Australian  Research Council   project No. DP140103925
  and   the Russian Science Foundation grant 14-42-00047 at Lebedev Institute.
  The research of AAT  at KITP  was also supported  in part by the National Science
Foundation under grant  No. NSF PHY11-25915.
AAT  also  thanks the Galileo Galilei Institute for Theoretical Physics for the
hospitality and the INFN for partial support during the completion of this work.

\def \C {{\cal C}}
\def \kk {\C}

\appendix

\section{Alternative  computational scheme for free energy and $C_T$  \\ of
 3-derivative conformal spinor   field} \la{app1}
%%%%%%%%%%%%%%%%%%%%%%%%%%%%%%%%%%%%

Let us  start with the standard  fermion  case and compute  the  corresponding free  energy 
without using  the proper time representation  for the log factor in \rf{63}    and 
doing the  sum  over $n$   first   and the integral over  $\l$ last. 
The sum over $n$  requires a  
 regularization prescription   and we shall  adopt   the same one  as  used, e.g., in 
   \cite{Klebanov:2011uf}\foot{This 
relation   may  be 
 justified, e.g.,  by     first taking  the derivative  over 
 $\mm$,  then doing the  convergent sum  using 
 $\sum^{\infty}_{n=-\infty}   {1 \ov [( n+\gamma)^{2}+  q^2 \mm^{2}]}
 = { \pi \sinh (2 \pi q\, \mm ) \ov  q\, \mm\,   [ \cosh(2\,\pi\,q\, \mm )- \cos(2\,\pi\,\gamma)]}$, 
 and finally  integrating  back   over $\mm$   %with  a particular integration constant to match \rf{4.8}  (and 
  (assuming also   that    $\sum^{\infty}_{n=-\infty} c =0$). 
  Note that the   choice of  integration constants  or  regularization involved   
  %AT28
  may break the  formal  invariance  under the  integer 
  shifts of    $\gamma$. }
\be
\la{4.23}
\sum^{\infty}_{n=-\infty}\log\big[\frac{(n+\gamma)^{2}}{q^{2}}+\mm^{2}\big] \Big|_{\rm reg} = 
\log\big[2\,\cosh(2\,\pi\,q\, \mm )-2\,\cos(2\,\pi\,\gamma)\big] \ . 
\ee
%A23
It is important to stress   that  because of the regularization involved 
this relation directly  applies  for $\gamma <1$    (and $q^2 \mm^2 > -1$ if $\g=0$); 
 the expressions found using \rf{4.23}  with parameters 
 outside that range should be defined by an analytic  continuation. 
The case of   half-integer   summation  in \rf{63}  corresponds to the legitimate values  $\g=\ha$, 
$\mm^2=\l \geq 0$.  
 As a result, we get 
\be\la{65}
\mc F_{q} = -\ha n_{f}\,{\VV_{\mathbb H^{d-1}}}
\int_{0}^{\infty}d\mu_{\frac{1}{2}, d-1}(\lambda)\,\log\big[2\,\cosh(2\,\pi\,q\,\sqrt\lambda)+2\big]\ .
\ee
The integral over $\l$ is   divergent at large  $\l$;  omitting the  power divergent part 
 proportional to $q$  one reproduces  the same  $d=4$ and $d=6$   expressions   as in \rf{5757} and \rf{7575}.  The second $q$-derivative  of $\F_q$  is always   finite 
 and   using  (\ref{3.3}), we  then   reproduce  from \rf{65}   the 
 standard  result   for  $C_{T,d}(\psi) = \ha n_{f}\, d$ \ 
   given in  (\ref{1.16}).

In the case of the 3-derivative  spinor  we may use the factorized  expression \rf{6.12}  for the square of its  kinetic operator   leading to the following  expression for the free energy  that  generalizes  \rf{63}
\ba
 \F_{q} &=-\text{tr}\,\log(i\,\OO^{(3)})=
 -\ha n_{f}\,{\VV_{\mathbb H^{d-1}}}
\int_{0}^{\infty}d\mu_{\frac{1}{2}, d-1}(\lambda)\,   K(\l, q)  \ , \la{0063}
\\
   K(\l, q) & \equiv   \sum_{n\in\mathbb Z+\frac{1}{2}}  \log \Big(
\big[\frac{n^{2}}{q^{2}}+\lambda\big]\,
\big[\frac{(n+q)^{2}}{q^{2}}+\lambda\big]\,
\big[\frac{(n-q)^{2}}{q^{2}}+\lambda\big]\Big) \ . \la{6300}
\end{align}
Computing  the sum in $K(\l, q)$ using  the prescription \rf{4.23} 
(with $\g$ equal to $\ha,\  q+\ha, \ q -\ha $)  we find the following analog of \rf{65} 
\begin{align}
\mc F_{q} =& -\ha n_{f}\,{\VV_{\mathbb H^{d-1}}}
\int_{0}^{\infty}d\mu_{\frac{1}{2},d-1}(\lambda)\, \Big(\log\big[2\,\cosh(2\,\pi\,q\,\sqrt\lambda)+2\big] \notag \\
&\  \ \  \qquad\ \ \  \  \ \  \qquad \qquad \  \ \  \qquad \ \ \ 
+
2\,\log\big[2\,\cosh(2\,\pi\,q\,\sqrt\lambda)+2\,\cos(2\,\pi\,q)\big]  \la{614}
\Big)\ .
\end{align}
%where the spectral measure was given in \rf{6.3}. 
Taking  the  second derivative   of \rf{614}   over $q$   at $q=1$   and  computing the 
resulting  finite integral  over $\l$   we find, 
 according to \rf{3.3},   
    the following  expression  for  the $C_T$ coefficient 
 for the 3-derivative  conformal fermion   in $d$ dimensions  (with  $n_f=2$ in $d=4$ and $d=6$)
\ba\la{615}
\te C_{T,d}(\psi^{(3)}) =    - n_{f} \, \frac{d(5\,d+11)}{2(d-1)}  \ , \ \ \ \ {\rm  i.e.} \ \ \ \ 
%= - \frac{5\,d+11}{d-1} \ C_{T,d}(\psi)  \, \ .
 \te  C_{T,4}  (\psi^{(3)})   = - \frac{124}{3} \ , \ \ \ \  \ 
 C_{T,6}(\psi^{(3)}) = - \frac{246}{5} \ .
\end{align} 
Remarkably, the  $d=6$  value is  precisely  the one in \rf{1180}   corresponding to $\xibt$ in \rf{1.8}.
However,  a warning   sign is  that  the $d=4$   value disagrees with the 
correct one $C_{T,4}= -{4\ov 3}$  in \rf{811} 
corresponding to $\cc= -{1\ov 120}$   found   by  direct  computation  in 
 \ci{Fradkin:1981jc,Fradkin:1985am}.
 
This  suggests   some problem    with  the above computation. Indeed, 
while the representation  \rf{65}   for    the free  energy  of  the standard   fermion is true  for any $q$, 
the expression \rf{614}  that was obtained using \rf{4.23}   with $\g = q  \pm  \ha$
is formally  valid only  for   $ 0 \leq  |q| < \ha$.  It cannot thus   be
 differentiated directly at $q=1$  and  this is the reason   why the resulting values  of $C_T \sim \F''_1$ or \rf{615}   are not correct.

 The  correct   procedure is to  first evaluate \rf{614}   for  $ 0 \leq  |q| < \ha$, then analytically extend the resulting  expression for $\F_q$ to all  values of $q$  and finally  differentiate  it over $q$ 
 obtaining, in particular,  the corresponding  \renyi entropy \rf{3.2} 
 and $C_T\sim \F''_1$  in \rf{3.3}. 
 The results  will then agree  with \rf{81},\rf{811} and \rf{y20},\rf{y39} 
 found  using  the  heat-kernel regularization  approach used  in the main text.

 %AT28
 To see this explicitly   let us note  that 
 the $+ 2\cos(2\,\pi\,q)= -2 \cos [ 2 \pi (q\pm \ha)]$  term  in \rf{614}  
originated  from the $\g= q \pm \ha$  shifts  in \rf{4.23}. To  make  the use of \rf{4.23}     legitimate  we  may   first formally   replace  this  shift  by $\g=  {q\ov k} \pm \ha$,  evaluate \rf{614}   for 
 $k>2 $  and then   analytically  continue  $k\to 1$   in the final result.  
   Replacing  $\cos(2\,\pi\,q) \to \cos(2\,\pi\,{q\ov k})$  
    in \rf{614}  we find   after   computing  the second derivative of \rf{614} at $q=1$ 
     in $d=4,\ 6$ for $k > 2 $  (see \rf{3346}) 
   \be \la{a777}
\te    C_{T,4} (\psi^{(3)}) = 40 \F''_1 = ( 6 - {20 \ov 3 k^2} )\, n_f \ , \qquad 
 C_{T,6}(\psi^{(3)})  = -252  \F''_1 =   (9 -\frac{161}{10k^2}+  \frac{7}{2k^4})\, n_f \ .
   \ee
    These  expressions indeed  reproduce the correct values $C_{T,4} = - { 2 \ov 3 } n_f$  in \rf{811} and   
    $C_{T,6} = - { 18 \ov 5 } n_f$  in \rf{811}   after the 
    analytic continuation to $k = 1$.\foot{Note  that for $k\to \infty$  the 
    expressions in \rf{a777} become  3 times the   standard  fermion  values in \rf{577}  and \rf{751}.
    The reason for this  is that in this limit   the $\pm q\to \pm {q\ov k}$ shifts of $n$  in \rf{6300}  disappear 
    and we get   the 3rd  power of the standard fermion expression under  the log.}

%%%%%%%%%%%%%%%%%%%%%%%%%%%%%%%%%%%%%%%%%%%%

\section{2-derivative  non-gauge conformal vector} %     without  gauge invariance}
\la{app:erd}

Here we shall    follow \cite{Erdmenger:1997gy,Osborn:2015rna}   and 
  consider   a  non-unitary   theory   described  by 
    2-derivative vector field   with conformal  but   not gauge-invariant  
     action for $d \not=4$.\foot{Partition function of a similar  2-derivative  
  spin 2 theory   was  discussed in \cite{Beccaria:2015vaa}.} 
The  corresponding Weyl-invariant  curved space action is 
\begin{align}
\la{B.1}\!\!\!
I= -\int d^{d}x\,\sqrt{g} \,  \Big[ \te D^{\m }V^{\n}\,D_{\m }V^{}_{\n}-\frac{4}{d}\,(
D^{\m }V^{}_{\m })^{2}+\frac{2}{d-2}\,R^{\mu\nu}\,V^{}_{\mu}\,V^{}_{\nu}
+\frac{d\,(d-4)}{4\,(d-1)(d-2)}\,R\,V^{\,\mu}V_{\mu}\Big].
\end{align}
It   is  equivalent to the standard Maxwell action  for  $d=4$. 
The  corresponding $C_T$ coefficient   found in \cite{Osborn:2015rna} is 
\be\la{b2}
 %d\neq 4:  \quad 
 C_{T,d\not=4}(V^{(2)}) = \tfrac{d^{2}}{d-1}\  ,  \ \ \qquad  \ \ \  \qquad
 C_{T,4}( V^{(2)}) = 16 \ .
\ee
%while the $d=4$ value is that of the standard Maxwell  theory,  $C_{T,4}( V^{(2)}) = 16$. 
Specifying to the case of the  $S^{1}\times \mathbb H^{d-1}$ background   and 
 separating  $V_{0}^{}$ and 
$V_{i}^{}= V^{\perp}_{i}+\del_{i}\,\varphi$  components  as in \rf{522} 
we find that the  corresponding partition function has two contributions: one  from  ${V}^{\perp}_i$
%with no shifts, 
and one  corresponding to  the $\partial^{4}$ conformal scalar in $d\neq 4$. The scalar  
part  is absent in $d=4$ due to gauge   invariance  that is   then present in \rf{B.1} (cf.  section 4.1   and \cite{Beccaria:2014jxa}). 

From (\ref{B.1})  we get  the following mixed   Lagrangian  for $\chi\equiv V_{0}^{},\ \varphi$ and   $V^\perp_i$ 
\begin{align}
\mathscr L = &\te  (D_{\mu}\chi)^{2}+(D_{\mu}V^{\perp}_{i}+D_{\mu}D_{i}\varphi)^{2}
-\frac{4}{d}\,(\partial_{0}\chi+\bm{D}^{2}\varphi)^{2}\notag \\
& \te  -2\,(V^{\perp}_{i}+D_{i}\varphi)^{2}
-\frac{d(d-4)}{4}\,\big[\chi^{2}+(V^{\perp}_{i}+D_{i}\varphi)^{2}\big]  
\equiv  \mathscr L(\chi, \varphi)+ \mathscr L (\bm{V}^{(2)}_\perp) \ , \la{b33}
\end{align}
where % (cf. \rf{555}  for $s=1$) 
\be \la{b44}
 \mathscr L(\bm{V}^{(2)}_\perp) = V^{\perp}_i \big( - \del_0^2  + \bm{\Delta}_1   \big) V^{\perp}_i
 \ , \qquad \qquad \bm{\Delta}_1 =  - \bm{D}^2 - \tfrac{(d-2)^2}{4} -1 \ . 
 \ee
%%%%%%%%%%%%%%%%%%%
Using that on  $\Ss^1\times \mathbb H^{d-1}$  we have 
$
D_{i}D^{2}D_{i}\varphi = \partial_{0}^{2}\bm{D}^{2}+(\bm{D}^{2})^{2}-(d-2)\bm{D}^{2},
$ 
we obtain  for the scalar part of \rf{b33} 
\begin{align}
\mathscr L(\chi,\varphi)&=\te \chi\,\big[\-    \frac{d-4}{d} \partial_{0}^{2}-\bm{D}^{2}-\frac{d(d-4)}{4}\big]\,\chi
+\frac{8}{d}\,\chi\partial_{0}\bm{D}^{2}\,\varphi
+\varphi\, \bm{D}^{2}\, \big[
\partial_{0}^{2}+\frac{d-4}{d}\,\bm{D}^{2}
+\frac{(d-4)^{2}}{4}
\big]\,\varphi.
\end{align}
%%%%%%%%%%%%%%%%%%%
Integrating over $\chi$ and $\varphi$ in the path integral,   their  contribution 
can be represented   in terms of   the determinant 
of the  following    6-order   scalar operator 
\be 
\OO^{(6)}
= \te -\frac{d-4}{d}\,\bm{D}^{2}\,\big[\partial_{0}^{4}+\frac{d^{2}-4d+8}{2}\, 
\partial_{0}^{2}+\big(2\partial_{0}^2+\frac{d(d-4)}{2}\big)\,\bm{D}^{2}+(\bm{D}^{2})^{2}\big].
\ee
The  determinant of the $\bm{D}^{2}$  factor cancels   against  the Jacobian of the change of variables $V_{i}^{}\to  V^{\perp}_{i}+\del_{i}\,\varphi$, 
while  the remaining 4-order  scalar operator is equivalent to the  conformal $\del^4$ one  
which factorizes as in \rf{4.21}  with the eigenvalues  given in \rf{417}. 

As a result, we find that  % (see  \rf{444}
\begin{align}
 C_{T,4}(V^{(2)}) = C_{T,4}(\bm V^{(2)}_{\perp}) \ ,   \qquad \qquad  %d\neq 4: \quad  
  C_{T,d\not=4}(V^{(2)}) = C_{T,d}(\bm V^{(2)}_{\perp})+C_{T,d}(\varphi^{(4)}) 
  %   \te  =  &\te \frac{d^{2}}{d-1}-\big[-\frac{2d(d+4)}{(d-1)(d-2)}\big]
%= \frac{d(d^{2}+8)}{(d-1)(d-2)}
 \ . \la{b66}
\end{align}
In view of the expression \rf{444}   for $C_{T,d}(\varphi^{(4)}) $,   to 
  match \rf{b2}   we should thus get 
\begin{align}
\la{B.5}
\!\!\!\! C_{T,4}(\bm V^{(2)}_{\perp}) = 16\  ,  \ \qquad \qquad 
%d\neq 4: \quad 
 \te C_{T,d\not=4}(\bm V^{(2)}_{\perp}) % = \frac{d^{2}}{d-1}-\big[-\frac{2d(d+4)}{(d-1)(d-2)}\big]
= \frac{d(d^{2}+8)}{(d-1)(d-2)}. \qquad \qquad 
\end{align}
The transverse  spatial  vector part of the free energy  that follows from \rf{b44}
is  given by (see \rf{x2}  for $s=1$, \rf{5.9}  and \rf{4.23})\foot{Here
%AT28
 we  use \rf{4.23} 
with $\gamma=0$ so the result is equivalent  to the one 
in the heat kernel approach  used in the main text.}
\begin{align}
\mc F_{q}(\bm V^{(2)}_{\perp}) &= \ha {\VV_{\mathbb H^{d-1}}}
\sum^{\infty}_{n=-\infty}\int_{0}^{\infty}d\mu_{1,d-1}(\lambda)\,\log\big(\frac{n^{2}}{q^{2}}
+\lambda\big) \notag \\
&= \VV_{\mathbb H^{d-1}}
\int_{0}^{\infty}d\mu_{1, d-1}(\lambda)\,\log\big[2\,\sinh(\pi\,q\,\sqrt\lambda)\big]\ .\la{b9}
\end{align}
Computing the  corresponding  $C_{T}$   according to \rf{3.3}  we find 
\be
C_{T,4}(\bm V^{(2)}_{\perp}) = 16, \qquad
C_{T,6}(\bm V^{(2)}_{\perp}) = \tfrac{66}{5}, \qquad
C_{T,8}(\bm V^{(2)}_{\perp}) = \tfrac{96}{7}, \ \  \ \ \  etc., 
\ee
in agreement with (\ref{B.5}).
This  provides  an  alternative  derivation of \rf{b2}. 

%\newpage

\section{Conformal  anomalies  of  general  higher derivative  short   \\
 superconformal  6d vector multiplets}
% from \\ Seeley -   DeWitt coefficients}
\la{app:Vp}

The calculation of the full 6d conformal anomaly of the $V^{(1,0)}$ multiplet
from the Seeley-DeWitt coefficients on an Einstein background 
presented in  section \ref{sec:seeley}  may be generalized to other (1,0) superconformal vector multiplets.
$V^{(1,0)}\equiv V^{(1,0)}_{p=2}$ is  the lowest  member of a family of 
multiplets $V^{(1,0)}_p$ ($p=2,3,4,\dots$)  that contain  scalars, spinors  and   vectors
with $p$-dependent higher-derivative   kinetic terms.  

In terms of $OSp(2,6|2)$ representations 
\cite{Minwalla:1997ka,Dobrev:2002dt,Bhattacharya:2008zy} the hypermultiplet $S^{(1,0)}$
is a doubleton ultra-short representation \cite{Ferrara:2000xg}.
New (possibly massive) conformal representations are 
obtained by tensoring $p$ copies of $S^{(1,0)}$. The resulting multiplets $V^{(1,0)}_{p}$
are  short with  the maximal spin equal to 1. The  structure of these multiplets 
 \cite{Gimon:1999yu}  is   shown in Table \ref{tab:Vp}.
 %%%%%%%%%%%%%%%%%%%%%%%%%%%%%%%%%%
\begin{table}[H]
\be
\def\arraystretch{1.3}
\begin{array}{cccc}
\toprule
\text{field} & SO(6) & SU(2)_{R} & \Delta \\
\midrule
 \varphi & (0,0,0) & \mathbf{p+1} & 2\,p \\
 \psi^{+} & (\tfrac{1}{2}, \tfrac{1}{2}, \tfrac{1}{2}) &  \mathbf{p}   & 2\,p+\tfrac{1}{2} \\
 V_{\m} & (1,0,0) & \mathbf{p-1} & 2\,p+1 \\
  \psi^{-} & (\tfrac{1}{2}, \tfrac{1}{2}, -\tfrac{1}{2}) &  \mathbf{p-2} & 2\,p+\tfrac{3}{2} \\
 \varphi' & (0,0,0) & \mathbf{p-3} & 2\,p +2\\
 \bottomrule
\end{array}\notag
\ee
%%%%%%%%%%%%%%
\caption{Short multiplets $V^{(1,0)}_{p}$ of $OSp(2,6|2)$ that  appear in 
tensor product of  $p$   copies of $(1,0)$ doubleton  (hypermultiplet) representation.
 }
\label{tab:Vp}
\end{table}
%%%%%%%%%%%%%%%%%%%%%%%%%%%%%%%%%%%%%
Here $\Delta$ is the scaling dimension
of the conformal group $SO(2,6)$ related to the canonical dimension of the corresponding 6d field $\Phi$ 
  by $\dim\Phi = 6-\Delta$. We indicated
   also  the $SU(2)$  R-symmetry representations.  $\psi^\pm$ are positive/negative  chirality MW   spinors while  $\vp$ and $\vp'$ are scalars. 
  The vector $V_\m$   has conformal but  not gauge invariant   action for $p >2$.
 From the canonical  dimensions   one can determine the number of derivatives in 
 kinetic terms in the corresponding 6d Lagrangian that  has  the following  schematic form 
  \be \la{Vp-31} 
\begin{split}
\mc L=& \varphi\,\Box^{2p-3}\, \varphi+\overline\psi^{+}\slashed{\partial}^{4p-5}\psi^{+}
+V_{\m}\,\Box^{2p-2}\,V_{\m}+\overline\psi^{-}\slashed{\partial}^{4p-3}\psi^{-}+
\varphi'\,\Box^{2p-1}\,\varphi'  \ , 
\end{split}
\ee
where fields transform under $SU(2)_{R}$ according to Table~\ref{tab:Vp}. 

The higher derivative  operators in \rf{Vp-31}   should have a covariant and 
Weyl-invariant    generalization to curved background.
 Remarkably, just as for the  $p=2$  case discussed in section 6
  all  these operators  for conformal
fields in Table \ref{tab:Vp}  factorize  on an  Einstein space background.
Let us denote by $\Phi^{(n)}$ a field with $n$ derivatives in the 
kinetic term.
For higher derivative conformal scalar $\varphi^{(2n)}$  operators (GJMS operators)
%General factorization of (so-called GJMS) 
the  factorization on an Einstein space  reads %is known 
\cite{Gover:2005mn}    %, and the conformal operator reads
\be
\prod_{k=1}^{n}\Delta_{0}\big(6-k(k-1)\big) \ , 
\ee
where $\Delta_0(\k)$ is the scalar Laplacian  defined in \rf{PP}.
Similarly, for the $(2n+1)$-derivative spinor $\psi^{(2\,n+1)}$, the square of the 
corresponding  conformal operator  factorizes as 
%of Dirac operator is in general
\cite{fischmann2015conformal} 
\be
\Delta_{1\ov 2 }(\tfrac{15}{2})\, \prod_{k=1}^{n}\big[\Delta_{1\ov 2 }(\tfrac{15-2\,k^{2}}{2})\big]^{2} \ , 
\ee
which generalizes the $n=1$   expression in \rf{P4}. 
Finally, for the  vectors $V^{(2n)}$, with $n>2$, the 
factorization on a general Einstein  space looks the same  as on  the  6-sphere
and can be found  from  (A.17) of \cite{Beccaria:2015uta} for the massive conformal
representation $[\Delta, \bm{h}]=[3+n, (1,0,0)]$. %Just as an illustration, 
The  corresponding partition function   can be written as  (cf. \rf{P2})
\be
\la{X4}
Z(V^{(2n)}) = \bigg[\frac{\Delta_{0}\big(-(n+3)(n-2)\big)\ \prod_{k=1}^{n}\Delta_{0}\big(-(k+1)(k-2)\big)}
{\prod_{k=1}^{n+1}\Delta_{0}\big(-(k+2)(k-3)\big)\ \prod_{k=1}^{n}\Delta_{1}\big(7+k-k^{2}\big)}\bigg]^{1/2}\ .
\ee
%where $\Delta_1(\k)$ 
These factorizations  into  2nd order Laplacians  allow us to  compute the corresponding 
conformal anomalies using the same method as  in 
 the  $V^{(1,0)}$ (i.e. $p=2$) case in section 6. 
%The existence of a curved space action for fields with arbitrarily high derivative order
%is doubtful and certainly false already in scalar case. 
%\red{Should we just say that the calculation is
%formal or argue that it may exist on Ricci flat ? } }
%From the explicit factorizations of conformal operators and 
Using the expressions for the 
Seeley - DeWitt coefficients $b_{6}$  in \rf{P.1},\rf{P.2},\rf{P.3}  we
 can compute
the anomalies 
of the  fields in the $V_{p}^{(1,0)}$ multiplet  with the results  summarized  below\footnote{Our discussion is  formal as curved-space higher-derivative operators in a given dimension 
(here $d=6$)  may exist only to some critical order (as is well 
known in the scalar GJMS case).}
\begin{align}
%\underline{\bm{\varphi^{(2n)}}} & \\
\varphi^{(2n)}: \ \ \ \  &7!\,{\rm a} = -\tfrac{1}{144} n^3 (3 n^4-21 n^2+28), \qquad 7!\,{\rm c_{1}} = -\tfrac{2}{9} n (3 n^2-5) (3 n^4-16 n^2-8),\no\\
&7!\,{\rm c_{2}} = -\tfrac{1}{18} n (9 n^6-63 n^4+112 n^2-88),\qquad 
7!\,{\rm c_{3}} = \tfrac{1}{6} n (n^6-7 n^4+18)\ , \no 
\\
%\underline{\bm{\psi^{(2n+1)}}} & \\
\psi^{(2n+1)}: \ \ \ \  &7!\,{\rm a} = \tfrac{1}{288} (2 n+1) (12 n^6+36 n^5-102 n^4-264 n^3+244 n^2+382
   n-191),\notag \\
&7!\,{\rm c_{1}} =\tfrac{2}{9} (2 n+1) (18 n^6+54 n^5-153 n^4-396 n^3+415 n^2+622
   n-336),\notag \\
&7!\,{\rm c_{2}} = \tfrac{1}{18} (2 n+1) (18 n^6+54 n^5-153 n^4-396 n^3+317 n^2+524
   n-144),\no\\
&7!\,{\rm c_{3}} = -\tfrac{1}{6} (2 n+1) (2 n^6+6 n^5-17 n^4-44 n^3+57 n^2+80
   n-60) ,\\
%\underline{\bm{V_{m}^{(2n)}}} & \\
V_{m}^{(2n)}: \ \ \ \  &7!\,{\rm a} = -\tfrac{1}{8} n^3 (n^4-14 n^2+21),\qquad 7!\,{\rm c_{1}} = -\tfrac{4}{3} n (9 n^6-126 n^4+231 n^2-2),\notag \\
&7!\,{\rm c_{2}} = -\tfrac{1}{3} n (9 n^6-126 n^4+147 n^2+80),\qquad 7!\,{\rm c_{3}} = n (n^6-14 n^4+35 n^2-10).\no
\end{align}
The total results  for the multiplet $V_{p}^{(1,0)} $  are  then  
\begin{align} V_{p}^{(1,0)}: \ \ \ \ 
&7!\,{\rm a}= 70\,(p-1)^{4}-35\,(p-1)^{2}+\tfrac{77}{48}, \no\\
&7!\,{\rm c}_{1}= 6720\,(p-1)^{4}-3920\,(p-1)^{2}+\tfrac{560}{3},\la{c11}  \\
&7!\,{\rm c}_{2}= 1680(p-1)^{4}-700\,(p-1)^{2}+\tfrac{28}{3}, \no \\
&7!\,{\rm c}_{3}= -560(p-1)^{4}+420(p-1)^{2}-28 \ . \no 
\end{align}
The expressions in \rf{c11} are in  perfect agreement  with 
\rf{1.5},\rf{1.2},\rf{1.7} with \rf{188}  as one can see using that 
for general $p$ the coefficients in the 
%The last line is in full agreement with the predictions obtained using $\xi=-\frac{8}{9}$ and the 
anomaly polynomial are \cite{Beccaria:2015ypa}
\be\la{cap}
(\alpha,\beta,\gamma,\delta) = \big(-(p-1)^{4}, -\tfrac{1}{2}(p-1)^{2}, -\tfrac{7}{240}, \, \tfrac{1}{60}\, \big). 
\ee
Remarkably, these expressions 
continue to hold also  for $p=2$ where (\ref{X4})  should  be replaced
by (\ref{P2}).\footnote{This agreement  is not accidental. For $p=2$ the vector $V^{(4)}$
has  gauge invariant action   (it is described by the 
conformal representation $[5,(1,0,0)]-[6,(0,0,0)]$ where the subtraction takes into account the gauge invariance),  while 
 the scalar $\varphi'$ is absent in  the multiplet $V^{(1,0)}$. In the general 
 expression   for  the partition function  for the multiplet $V^{(1,0)}_p$
 continued formally  to  $p=2$ the field $\varphi'$
 enters   effectively with a negative multiplicity  and thus 
contributes precisely  like the ghost  scalar factor  in   \rf{P2}.
}
%
%The above  conformal anomaly coefficients for  the $p=3$ multiplet 
%have the following generalization to any $p\geq 2$ 
%%Indeed, replacing $p=3$ in the general expressions (of course $c_{1}-2c_{2}+6c_{3}=0$)
%\begin{align}
%7!\,{\rm a}(V_{p}^{(1,0)}) &= \tfrac{1}{72}\,(p-1)^{4}-\tfrac{1}{144}\,(p-1)^{2}+\tfrac{11}{34560}, \notag \\
%7!\,{\rm c}_{1}(V_{p}^{(1,0)}) &= 6720\,(p-1)^{4}-3920\,(p-1)^{2}+\tfrac{560}{3}, \notag \\
%7!\,{\rm c}_{2}(V_{p}^{(1,0)}) &= 1680\,(p-1)^{4}-700\,(p-1)^{2}+\tfrac{28}{3}, \notag \\
%7!\,{\rm c}_{3}(V_{p}^{(1,0)}) &= -560\,(p-1)^{4}+420\,(p-1)^{2}-28,
%\end{align}
%%we match the last line of (\ref{X1}).
%which are again in agreement  with \rf{1.5},\rf{1.2},\rf{1.7},\rf{188} and \rf{cap}.

%%%%%%%%%%%%%%%%%%%%%%%%%%%%%%%%%%%%%
\bibliography{BT-Biblio}
\bibliographystyle{JHEP}

\end{document}